\documentstyle[sprocl]{article}
\input{psfig}
\begin{document}
\bibliographystyle{unsrt}    
\def\Journal#1#2#3#4{{#1} {\bf #2}, #3 (#4)}
\def\NCA{\em Nuovo Cimento}
\def\NIM{\em Nucl. Instrum. Methods}
\def\NIMA{{\em Nucl. Instrum. Methods} A}
\def\NPB{{\em Nucl. Phys.} B}
\def\PLB{{\em Phys. Lett.}  B}
\def\PRL{\em Phys. Rev. Lett.}
\def\PRD{{\em Phys. Rev.} D}
\def\ZPC{{\em Z. Phys.} C}
\newcommand{\spur}[1]{\not\! #1 \,}
\def\st{\scriptstyle}
\def\sst{\scriptscriptstyle}
\def\mco{\multicolumn}
\def\epp{\epsilon^{\prime}}
\def\vep{\varepsilon}
\def\ra{\rightarrow}
\def\ppg{\pi^+\pi^-\gamma}
\def\vp{{\bf p}}
\def\ko{K^0}
\def\kb{\bar{K^0}}
\def\al{\alpha}
\def\ab{\bar{\alpha}}
\def\be{\begin{equation}}
\def\ee{\end{equation}}
\def\bea{\begin{eqnarray}}
\def\eea{\end{eqnarray}}
\def\CPbar{\hbox{{\rm CP}\hskip-1.80em{/}}}







\newpage
\title{WEAK DECAYS OF HEAVY QUARKS}

\author{ FULVIA DE FAZIO }
\address{Center for Particle Theory, University of Durham, \\
Durham, DH1 3LE, U.K.} 

\maketitle\abstracts{We review some aspects of weak decays of hadrons
containing one heavy quark. The main emphasis is on $B$ physics, in
particular in the framework of the Heavy Quark Effective Theory.}


\vspace{1cm}

\tableofcontents
\newpage

\section{Motivations}
In the last decades we have witnessed an impressive progress in our
understanding of elementary particle physics phenomena. On the one hand,
more and more powerful experimental tools became available, on the other a
deeper insight in the theoretical understanding of elementary dynamics
has been
gained. We have reached a point where the Standard Model describing strong
and electroweak interactions of elementary particles has obtained
unforeseen confirmations and, simultaneously, the many questions 
which still
remain open  push us to look beyond it. This is the frontier of
particle physics, to which all of  this book is devoted. 

Among the many unanswered questions, the mechanism of
electroweak symmetry breaking plays a primary role 
-- the Higgs particle which enters in the Standard
Model in the description of such a mechanism
has not yet been detected.
In the fermion sector, it is mostly unclear
as to  the observed hierarchy between 
the masses of the particles, in particular it is being 
debated whether or not we
can still believe that  neutrinos are massless;  
as far as quarks are concerned, problems already arise 
in trying to define their masses, due to the confining strength of QCD 
interactions in the long distance regime. 

Another  central question is  how do quarks
participate in weak decays: i.e.  the actual understanding of the
mixing pattern. This last point has especially attracted attention both
from experimentalists and theorists. The Standard Model describes such
mixing by introducing a unitary matrix, the
Cabibbo-Kobayashi-Maskawa (CKM) matrix. The complex nature of
this matrix is the only source of CP violation within the Standard
Model: many efforts are devoted to understanding 
whether this is sufficient
to describe all CP violating phenomena observed till now and all those
which will  hopefully be observed in the nearby future. This is considered
one of the most fertile grounds to reveal physics beyond the Standard
Model. The requirement of CP violating processes as one of the
necessary conditions to generate the observed baryon-antibaryon
asymmetry$\;$\cite{sak} makes this topic extremely appealing also from
the standpoint of cosmology.

Weak decays are concerned with all these and many
other left aside questions. The phenomenology of such decays is
impressively rich, and a huge number of data has been collected  at
the numerous experimental facilities operating in the past and at
present all over the world.
On the other hand,  most of these data come from hadronic processes,
and hence quantum chromodynamics (QCD) enters into  the game. 
QCD is for sure the less easily tractable theory in elementary particle
physics.  
While the perturbation theory has
provided us with a great predictive power in quantum electrodynamics
(QED), 
the limited realm of applicability of such an approach to the strong
interactions precludes to us the possibility to give equally precise
predictions in QCD.
When facing with weak decays of hadrons, we have to take into account that
these objects are  bound states of quarks and gluons, which is a highly
nonperturbative effect. Therefore, the interplay of both interactions should
always be considered, and this makes our task more difficult.

From this point of view a particularly favorable sector  is the one 
where
hadrons contain a single heavy quark.
The large scale is set by the heavy quark mass
$m_Q$ which gives us the possibility to exploit asymptotic freedom.
 Actually,
useful considerations can be derived when considering the limit $m_Q
\to \infty$. This is realized within the Heavy Quark Effective Theory
(HQET)$\;$\cite{isw}$^{\!-\,}$\cite{georgi} which
implements systematically new and old ideas on the dynamics of heavy quark
systems.\cite{shuryak,hqet_1}

The  decays of heavy flavored hadrons, in particular $B$ decays, will
presumably also show  a higher degree of CP violation than the
kaon system, where this phenomenon was first observed. This is why
new high luminosity machines ({\it $B$ factories}) 
have been built to study CP violation in $B$ decays. The 
possibility to observe a contribution of non-Standard Model
particles, whatever they are, in loop-induced rare $B$ decays, makes this
sector appealing from both theoretical and experimental sides.

These are the main motivations to study weak decays of heavy quarks, which
we shall review in the following. 
The first step will be the construction of the effective weak Hamiltonian
for $B$ decays. This is useful in many respects: it  allows to
discuss non leptonic and rare $B$ decays; it will be a guide in the
discussion of the Heavy Quark Effective Theory, the construction of 
which, however, follows a slightly different pattern. 
Next, we shall  briefly recall the formulation of the 
Heavy Quark Effective
Theory,
focusing on the aspects which will be relevant for the subsequent
analyses. A detailed construction of the Heavy Quark Theory is given
elsewhere in this book.\cite{uraltsev}
As an  application of the HQET formalism, we will discuss the
relations for leptonic
constants and semileptonic decay form factors. Very often in such cases,
the large $m_Q$ technique has been used in conjunction with QCD sum
rules,\cite{shifman} 
which are reviewed elsewhere.\cite{col}
 
The ${1 / m_Q}$ expansion has also proven to be
a very useful tool 
in the
computation of inclusive decays of heavy flavored hadrons. These have
attracted considerable interest due to the possibility of reducing the
hadronic uncertainty in the sum of exclusive modes and also because of
open questions concerning several
semileptonic $B$ decay modes and the ratios of  the beauty hadron
lifetimes. We shall survey  these problems in the considered
context.

Finally, we shall give some
conclusions, with a short comment on some unaddressed topics.

\section{Effective Weak Hamiltonian for $B$ decays}\label{hamilt}
In the Standard Model, weak decays of quarks and
leptons are mediated by the  vector bosons $W^\pm, ~Z^0$. The fact
that these particles are very heavy with respect to the energies usually
involved in weak decays of   hadrons, makes very convenient the use of an
effective theory in which these particles are explicitly integrated out
of the generating functional of the theory and therefore
no longer play  a
dynamical role. In particular, integrating out the $W$ boson leads to the
familiar pointlike four-fermion interaction  
originally introduced by Fermi in order to describe the nuclear $\beta$ decay.
The path which leads from the Standard Model Lagrangian to the Fermi one
is nothing but an example of the application of the operator product
expansion$\;$\cite{wilson} (OPE).
The virtue of using such a
description 
traces back to the interplay of strong and weak interactions.
The OPE allows us to write the amplitude relative to a certain
weak decay process as the matrix element of an effective Hamiltonian,
which is built as a sum of local operators weighted by suitable
coefficients (Wilson coefficients). In this context, the operators can be
viewed as effective vertices, and the coefficients as effective coupling
constants. Since this construction follows the procedure of integrating
out the $W$ boson, this particle does not appear in the
operators, but its effects are  contained in the coefficients. 
The expansion can roughly be viewed  as an expansion both in the inverse
powers of $M_W$ and  in the dimension of the operators,
the most important contribution coming from dimension six four-fermion
operators, since higher contributions are suppressed by powers of
$p^2/M_W^2$, $p$ being a typical momentum scale of the considered
process. In the case of the Fermi theory, it is easy to see what
has been just stated: the local four-fermion operator appearing in the
Lagrangian is completely independent on $M_W$; however, the Fermi
constant contains the dependence on it,
 $${{G_F \over
\sqrt{2}}={g^2\over 8M_W^2}}\,,$$ 
$g$ being the $SU(2)_L$ gauge constant.

Most importantly, the OPE succeeds in separating the long 
distance regime from the
short distance one. In fact, both the coefficients and the 
operators in the
OPE depend on a scale $\mu$. This is defined by the fact that we integrate
out all the particles with masses above this scale, therefore
the operators describe the physics below $\mu$,
while the coefficients take into account the physics above it, i.e. the
short distance regime. This is why they are usually referred to as ``short
distance coefficients."
However, the choice of $\mu$ is arbitrary, so that the dependence on it
should cancel between operators and coefficients. In this way we have
disentangled the nonperturbative ingredients of our process, i.e. the
matrix elements of the operators in the OPE, from the purely perturbative
ones, the Wilson coefficients. Therefore, a good choice of $\mu$ is
obtained requiring that the strong coupling constant is low enough to make
meaningful the perturbative calculation. In $B$ and $D$  decays a common
choice  is usually $\mu={\cal O}(m_b)$, $\mu={\cal O}(m_c)$, respectively. 
The Wilson coefficients are determined by requiring that, at a definite
scale, the amplitudes relevant to a process should coincide in the full
and in the effective theory, a procedure called {\it matching}.
The perturbative calculation typically shows the appearance of
large logarithms of the kind,
 $${\alpha_s(\mu) \ln
\left({M_W^2 \over \mu^2}\right)}\,,$$ so that, even when $\alpha_s(\mu)$
is a good expansion parameter, this  product is ${\cal O}(1)$, and
therefore the validity of the perturbative expansion is spoiled. The
powerful tools of the renormalization group help us  to sum up
these large logarithms, using what is called 
the {\it renormalization group
improved perturbation theory}. The leading term in this case stems from
the resummation of the terms 
$${\left[\alpha_s(\mu) \ln
\left({M_W^2 \over \mu^2}\right)\right]^n}$$ 
(the so-called leading log
approximation). In general, at order $m$ in this new expansion, one
resums
the terms 
$${\alpha_s(\mu)^m\left[\alpha_s(\mu) \ln   
\left({M_W^2 \over \mu^2}\right)\right]^n}\,.$$
Generally speaking, it is often insufficient to stop at the
leading log
approximation, and the next-to-leading order corrections should be included,
which are actually the truly ${\cal O}(\alpha_s)$ corrections with respect
to the leading term and which show many interesting features
hidden in the leading log approximation.\cite{buras} 

Keeping in mind the general steps necessary to build an effective
Hamiltonian, let us write explicitly the effective Hamiltonian describing
nonleptonic weak decays of $B$ mesons:
\be
H_{eff}^B={G_F \over \sqrt{2}} \sum_{U=u,c} V_{Ub}~
[C_1(\mu) ~Q_1^{Ub}+C_2(\mu) ~Q_2^{Ub}]+ {\mbox h.c.}
\label{heffb}
\ee
\noindent where we have omitted penguin operators and
\bea
Q_1^{Ub}&=&({\bar U}b)_{V-A}~[({\bar d}^\prime u)_{V-A}+({\bar s}^\prime
c)_{V-A}] \; , \nonumber \\
Q_2^{Ub}&=&({\bar U}u)_{V-A}({\bar d}^\prime b)_{V-A}+({\bar U}c)_{V-A}
({\bar s}^\prime b)_{V-A}\;\;, \label{q12}
\eea
\noindent with the short notation $({\bar q}_1 q_2)_{V-A}={\bar q}_1
\gamma_\mu (1-\gamma_5) q_2$ and $U=u,~c$. The primed fields are the
weak interacting
quarks, related to the mass eigenstates by the CKM mixing
\be
\left( \begin{array}{c} 
d^\prime \nonumber \\
s^\prime \nonumber \\
b^\prime \nonumber 
\end{array}\right )=
\left( \begin{array}{ccc}
V_{ud} & V_{us} & V_{ub} \nonumber \\
V_{cd} & V_{cs} & V_{cb} \nonumber \\
V_{td} & V_{ts} & V_{tb} \nonumber 
\end{array}\right )
\left( \begin{array}{c}
d \nonumber \\
s \nonumber \\ 
b \nonumber   
\end{array}\right ) \;\;.
\end{equation}
\noindent Let us observe that $Q_1^{Ub}$ corresponds to the same operator that
we would have found in the Fermi pointlike approximation. In such an
approximation, one has $C_1=1,~C_2=0$. QCD corrections have both the
effect of introducing a new operator, with the same flavor content, but a
different color structure, and of modifying the Wilson coefficients.
 
We will need the general features of effective theories 
recalled in this section, in order to understand the relation between 
HQET and QCD, as well as to stress the difference arising in this
situation where we just want to describe heavy quark dynamics and
therefore cannot completely integrate them out.

\section{Heavy Quark Effective Theory}\label{hqt}
The aim of this section is to give an intuitive picture of  heavy quark
symmetries and the basis of the systematic HQET formulation.\cite{neubpr}

Let us recall what is specifically meant by heavy quark. 
Except for the quark masses, QCD has a single
adjustable parameter,  $\Lambda_{QCD}$, which could be viewed as
the scale separating the strong coupling regime from the weak one. It
depends on the number of ``active" flavors at a given scale. 
It cannot be precisely determined, but, roughly
speaking, it can be considered as the inverse of the radius of a hadron,
where confinement effects act to give the bound state. For $R_{had}\simeq
1$~fm, it turns out that $\Lambda_{QCD}\simeq 200$~MeV. Therefore we can
classify heavy quarks as those with mass  larger than this
parameter. Since we expect that typical strong interactions inside the
hadron  involve exchange of gluons with a
virtuality  of order
$\Lambda_{QCD}$, this gives the heavy quark a special role in the hadron,
as
we shall see.  

We will therefore consider $c,~b,~t$ as heavy quarks, while
$u,~d,~s$ will be considered light. However, as well as the strange quark
cannot be  always considered light in the application of chiral
perturbation theory techniques, in the same sense the considerations which
are derived within HQET should be applied with great caution to charm.
Finally,  the top quark cannot be considered at all, since it will decay
before any bound state can be formed. Hence, strictly speaking, HQET is
a theory of $b$ decays. 

In order to describe qualitatively a system containing a heavy quark $Q$,
we will consider the explicit case of a heavy meson, and we will refer to
its light antiquark plus the cloud of gluons as the light degrees of
freedom. We would like to consider this system in the large heavy
quark mass limit, $m_Q \to \infty$. In this limit 
(which is is quite close to
reality) the chromomagnetic interaction between the heavy quark
spin $s_Q$ and the light degrees of freedom total angular momentum
$s_\ell$ dies off -- it 
is inversely proportional to the heavy quark mass (just like the
familiar quantum mechanical spin-orbit interaction).
Therefore such quantities are expected to decouple in the heavy quark
limit. Besides, in the QCD Lagrangian the dependence on the quark flavor
is only contained in the mass term, so we also expect that the
heavy quark flavor becomes irrelevant  for $m_Q \to \infty$.
This is the intuitive content of the, by now, well-known heavy quark spin
symmetry realized within HQET. This symmetry tells us 
that the light degrees of freedom in a heavy hadron are the same
independently of the heavy quark spin and flavor quantum numbers.  

One notes more than once the strict
analogy with atomic physics, observing that the flavor symmetry could be
compared  to the fact that different isotopes
of the same element have
identical chemical properties, while the spin symmetry translates in
atomic physics in the (almost) degeneracy of hyperfine levels.
These observations already point out a limit of the considered framework,
since they allow us to exploit the decoupling of the heavy quark in order
to work out relations among  various quantities involved in heavy hadron
decays; on the other hand, they do not allow us to say anything about the
light degrees of freedom.

Another result that can be intuitively understood is known in the literature
as the velocity superselection rule, which is due to Georgi.\cite{georgi}
The momentum of a heavy hadron $H_Q$  can be written as $p_H^\mu=m_H
~v^\mu$, while for the heavy quark: $p_Q^\mu=m_Q\; v^\mu +k^\mu$, where
$k$ is a residual momentum of ${\cal O}(\Lambda_{QCD})$. Since, on the
other hand $p_Q^\mu=m_Q\; v_Q^\mu$, one finds that, for  $m_Q \to
\infty$, $v_Q^\mu=v^\mu$, i.e. strong interactions conserve the heavy
quark velocity.

In order to give a systematic structure to these intuitive properties, we
would like to build up an effective field theory for large $m_Q$. However,
the procedure outlined in the previous paragraph is not very useful if we
only want to describe heavy hadron interactions. To use a familiar
example, if we were to describe
processes  where the $W$ boson appears as a final or initial state, we
could no more use the effective Fermi Lagrangian where the $W$ is no more
a dynamical field. For this reason, HQET is not obtained
integrating out the heavy quark, but only the ``small" components
of its spinor. Therefore,
we have to start from the QCD Lagrangian relative to the heavy quark 
\be
{\cal L}_Q={\bar Q}(i \spur D -m_Q)Q\;\; \label{qcdlagr}
\ee
\noindent where $D$ is the covariant derivative in the fundamental
representation,
and redefine the heavy quark spinor $Q$ as follows:
\be
Q(x)=e^{-im_Q v \cdot x}[h_v(x)+H_v(x)] \;\;,\label{q}
\ee
where $h_v$ and $H_v$ are defined by the application of the velocity
projectors,
\be
h_v={1 + \spur v \over 2}Q \qquad
H_v={1 - \spur v \over 2}Q  \;\;, \label{hv}
\ee
and therefore they satisfy:
${\spur v}h_v=h_v$, ${\spur v}H_v=-H_v$.
Substituting  Eq.~(\ref{hv}) in Eq.~(\ref{qcdlagr}) one can derive the
equation of motion for $H_v$,
\be
H_v={1 \over i v\cdot D +2
m_Q} i \spur D_\perp h_v \;,\label{eqmot}
\ee 
where $D_\perp^\mu=D^\mu-v \cdot D v^\mu$. Equation (\ref{eqmot})
 can be used to eliminate $H_v$ from the Lagrangian, leading to
the following one,  equivalent to ${\cal L}_Q$:
\be
{\cal L}={\bar h}_v i v \cdot D h_v+ {\bar h}_v i \spur D_\perp {1 \over i
v\cdot D +2
m_Q} i \spur D_\perp h_v \;.\label{inter}
\ee
The second term can now be expanded in the powers of $1/
m_Q$, yielding
\be
{\cal L}={\bar h}_v i v \cdot D h_v+{1 \over 2 m_Q} {\bar h}_v (i \spur
D_\perp)^2 h_v +{1 \over 2 m_Q}{\bar h}_v {g_s \sigma_{\alpha \beta}
G^{\alpha \beta} \over 2 } h_v + {\cal O} \left({1 \over  m_Q}\right)^2
\label{lag1m} \; .
\ee
The second term represents the  kinetic energy of the heavy quark
due to its residual momentum $k$. The last term stems from the
chromomagnetic coupling of the heavy quark spin to the gluon field. The
fact that this term appears only at the order $1/m_Q$
is the
origin of the spin symmetry. 
At the leading order in the heavy quark expansion, we simply have
\be
{\cal L}_{HQET}={\bar h}_v i v \cdot D h_v \label{lhqet} \;,
\ee
from which the following Feynman rule for the heavy quark
propagator can be derived:
\be
{i \over v \cdot k} { 1 + {\spur v} \over 2} \;\;, \label{prop}
\ee
\noindent as well as the one for the quark-gluon vertex,
\be
i ~g_s T^a v_\mu \;\;. \label{vertex}
\ee
\noindent $T^a$ are the $SU(3)$ generators, $T^a=\lambda^a/2$, where
$\lambda^a$ stand for the Gell-Mann matrices.
Again, we recognize that spin symmetry holds from the absence of
gamma matrices in Eq.~(\ref{vertex}); besides, the independence of
${\cal L}_{HQET}$ of $m_Q$ signals  the flavor symmetry. 
Most of the results that we will report in this chapter exploit
the symmetries of ${\cal L}_{HQET}$. On the other hand, the predictive
power of the approach
could be improved assessing the role of symmetry-breaking terms,
i.e.  including subleading terms in the heavy quark expansion. This can be
done  starting from the result in Eq.~(\ref{eqmot}), which gives us the
possibility to write each QCD operator as an expansion in the
powers of ${1
/  m_Q}$. In fact, expanding Eq.~(\ref{eqmot}), one can derive
the
expression for the heavy quark field $Q$ in QCD at 
the order ${1 /
m_Q}$,
\be
Q(x)=e^{-i m_Q v \cdot x} \left(1+{i {\spur D}_\perp \over 2 m_Q}
\right)  h_v
\label{q1m} \;,
\ee 
\noindent from which we can build any desired QCD operator, for example a 
current, at the considered order.  In practice  we are
interested in  the matrix elements of such operators, which we would
like to express by an expansion in ${1 /m_Q}$. 
However, 
the states on which the operators act still contain a dependence on $m_Q$,
since, at the next to leading order in the heavy quark
expansion, $h_v$ does not
 satisfy the same equation of motion stemming
from Eq.~(\ref{lhqet}).
The way out is to consider once
and  for all Eq.~(\ref{lhqet}) as the HQET
lagrangian; then $h_v$ exactly satisfies  the equation ${\spur 
v}h_v=h_v$. The subleading terms in Eq.~(\ref{lag1m}) are treated as
perturbations. In this
way, as in the usual spirit of the perturbation theory, the results obtained
in the presence of the perturbation can be expressed in terms of the
unperturbated quantities.

As a concrete example of how an asymptotic relation could be modified by
the inclusion of ${1 /m_Q}$ corrections, let us consider
the
heavy hadron masses. 
On the basis of spin symmetry, one can argue that hadrons differing only
in the orientation of the heavy quark spin should be degenerate in mass.
For example, this should hold in the case of the pseudoscalar and
vector heavy flavored mesons. In the charm case the experimental values
give: $m_{D^*}-m_D \simeq 140$~MeV, while for the beauty we get:  
$m_{B^*}-m_B \simeq 47$~MeV. This is not surprising, since we 
expect heavy quark symmetry to work better in the $b$ case.
However, we can go further by
 predicting the size of the corrections: if
these are ${\cal O}\left({1/ m_Q}\right)$, one should
have
$$m_{D^*}^2-m_D^2\simeq m_{B^*}^2-m_B^2\,.$$
 Experimentally this is quite well
confirmed,
 $$m_{D^*}^2-m_D^2\simeq 0.49~{\rm GeV}^2\,,\quad m_{B^*}^2-m_B^2\simeq
0.55~{\rm GeV}^2\,.$$
The previous considerations allow us to express the ${1/
m_Q}$
corrections in terms of the matrix elements of the perturbations appearing
in Eq.~(\ref{lag1m}). 
Let us define the expectation values of the two subleading terms in 
Eq.~(\ref{lag1m}) on a hadron $H_Q$ as follows: 
\bea
\mu_\pi^2&=&{1 \over 2 m_Q}~\langle  H_Q| {\bar h}_v (i \spur
D_\perp)^2 h_v| H_Q \rangle   \label{mupi} \; , \\
\mu_G^2 &=& {1 \over 2 m_Q}~\langle  H_Q|{\bar h}_v {g_s \sigma_{\alpha \beta}
G^{\alpha \beta} \over 2 } h_v |H_Q \rangle  
\label{mug}\;\;.
\eea
\noindent Notice that the states are
normalized to $2m_Q$, so that the previous quantities are independent of 
$m_Q$.
Using this notation, the mass of $H_Q$ can be written as
follows:
\be
M_{H_Q}=m_Q+{\bar \Lambda}+{\mu^2_\pi-\mu_G^2 \over 2 m_Q}+ {\cal O}
\left({1 \over  m_Q^2}\right) \label{hadmass} \;\;.
\ee
\noindent Since the chromomagnetic operator is responsible for the
spin symmetry breaking, and hence for the mass difference of the hadrons
which differ only in the orientation of the heavy quark spin, it is
possible to relate $\mu_G^2$ to such a difference. In particular, one can
write $\mu^2_G=-2\left[ J(J+1)-{3 \over 2} \right] \lambda_2$, where
$\lambda_2$, as well as ${\bar \Lambda}$ and $\mu^2_\pi$, does not depend
on $m_Q$.  From the measured mass splitting in the beauty mesons, where
 we expect our considerations to hold more 
accurately, it is possible to predict 
$$\lambda_2={1 \over
4}(M_{B^*}^2-M_B^2) \simeq 0.12~{\rm GeV}^2\,.$$

On the other hand, in the case of $B$ mesons, different controversial
determinations of $\mu^2_\pi$
exist in literature,\cite{mu2pi} mainly due to the difficulty of
extracting this
parameter from experimental data. Finally, ${\bar \Lambda}$ has the role
of the mass difference between the heavy hadron and the heavy quark in the
$m_Q \to \infty$ limit. It is of ${\cal O}(\Lambda_{QCD})$. 

Let us observe that the relation $Q(x)=e^{-i m_Q v \cdot x} h_v(x)$,
stemming from Eq.~(\ref{q1m}) in the heavy quark limit, {\it defines} the
heavy
quark mass $m_Q$.  This turns out  to be a physical parameter, given by
the relation $m_Q=M_{H_Q}-{\bar \Lambda}$, in the same limit. For example,
the form factors describing the decay $\Lambda_b \to \Lambda_c e
{\bar \nu}_e$ depend on ${\bar \Lambda}=M_{\Lambda_b}-m_b$;\cite{grin}
therefore, one could use  this decay to determine ${\bar \Lambda}$, so
that  the experimental value of $M_{\Lambda_b}$ would provide a
determination of $m_b$. In this sense, the mass appearing in
Eq.~(\ref{hadmass}) can be considered as the nonperturbative analog of the
pole mass, which, despite being a well defined quantity in perturbation
theory, suffers from renormalon ambiguities beyond it.\cite{renormalon}
We shall see  later in the discussion of inclusive decays that the
adoption of such definition leads to the absence of ${1 /
m_Q}$ corrections to the partonic prediction.

\subsection{Weak Decays of Heavy Mesons}
An important role is played by heavy quark spin-flavor symmetries when
considering weak decays of heavy mesons. Let us consider for example the
elastic scattering of a pseudoscalar meson $P(v)$ induced by an external
vector current coupled to the heavy quark. Since in the $m_Q \to \infty$
limit the light degrees of freedom are decoupled from the heavy quark, it
looks sensible to describe this process as if the current were acting only
on the heavy quark. We can write 
\be
\langle P(v^\prime)|{\bar h}_{v^\prime} \gamma^\mu
h_v|P(v)\rangle =\langle Q,v^\prime,s_Q^\prime|{\bar h}_v
\gamma^\mu h_v|Q,v,s_Q\rangle ~\langle L,s_\ell^\prime|L,s_\ell\rangle  \label{fact}
\ee
\noindent where $L$ represents the light degrees of freedom. The content
of Eq.~(\ref{fact}) amounts to a {\it factorization}, where the last term
represents the overlap of the light degrees of freedom.
If $v=v^\prime$, spin symmetry assures us that nothing is changed for the
light degrees of freedom: the overlap is simply
$\delta_{s_\ell,s_\ell^\prime}$. This holds also if the heavy quark in the
final state is a different one, thanks to the flavor symmetry. 

If $v \neq v^\prime$, the decoupling of
the light degrees of freedom still allows us to write such a 
factorization. However,
now the light degrees of freedom in the final state are in general changed
with respect to the  initial state and the overlap will be a new function
$\xi(v \cdot v^\prime)$  not depending on the spin and the flavor
of the heavy quark, and which is normalized to 1 for $v \cdot
v^\prime=1$,
since the zero component of the vector current is the
generator of the flavor symmetry. A detailed proof
of such normalization will be given later.
This function is usually referred to as the
Isgur-Wise universal form factor.\cite{isw,georgirev} Its universality is
apparent if we
consider that it is possible to describe the semileptonic decays $B \to
D$ or $B \to D^*$ in terms of this single function, with the  result of  a
great simplification of the theoretical study of these processes, which
are usually described by 2 and 4 form factors, respectively.
The possibility to give asymptotic relations between these form factors is
a quite appealing feature. As a matter of fact, one could consider such
relations as a test for models predicting these quantities or for the
outcome of nonperturbative techniques, such as lattice or QCD sum rules.
For example, Neubert has shown that not one of the commonly used quark
models proposed in the literature satisfied the asymptotic behaviour
predicted by the heavy quark symmetry.\cite{neubpr}

Interestingly enough, the argument that lead us to the introduction of the
Isgur-Wise function can be repeated in similar situations. 
Spin symmetry tells us that for each value of the light degrees of freedom
total angular momentum $s_\ell$ there are two degenerate states with total
spin: ${\vec J}={\vec s}_Q+{\vec s}_\ell$, according to the rules of
addition of angular momentum. These ideally degenerate states
will conveniently fit into doublets.
Besides, ${\vec s}_\ell$ can be written as the sum
of the light antiquark spin ${\vec s}_q$ plus the  orbital
angular momentum $\vec \ell$ of the light degrees of freedom with respect
to the heavy quark. Therefore, a given doublet could be identified by the
value of $s_\ell$ and the parity of the states $P=(-1)^{\ell+1}$.
If $\ell=0$, the two degenerate states form the doublet of pseudoscalar
and vector mesons $(P,P^*)$ (with $P=D,~B$); these have
$J^P_{s_\ell}=(0^-,1^-)_{1/2}$.
Therefore the Isgur-Wise function is the form factor describing the
decay of an element of this doublet to another element of the same
doublet. 

Let us  consider the case $\ell=1$. Now it could be either $s_\ell=1/2$
or $s_\ell=3/2$. The two corresponding doublets have positive parity; they
are: $J^P_{s_\ell}=(0^+,1^+)_{1/2}$, $J^P_{s_\ell}=(1^+,2^+)_{3/2}$. 
We will refer to elements of these doublets generically
as $(P_0,P_1^\prime)$ and $(P_1,P_2)$, respectively. 
The charmed $2^+_{3/2}$ state  has been experimentally observed
and denoted as the $D_2^*(2460)$ meson, with
$m_{D_2^*}=2458.9\pm 2.0~{\rm MeV}$, $\Gamma_{D_2^*}=23\pm 5~{\rm MeV}$ and
$m_{D_2^*}=2459\pm 4~{\rm MeV}$, $\Gamma_{D_2^*}=25^{+8}_{-7}~{\rm MeV}$ 
for the
neutral
and charged states, respectively.\cite{pdg}
The HQET state $1^+_{3/2}$ can be identified with $D_1(2420)$, with
$m_{D_1}=2422.2\pm 1.8$ MeV and
$\Gamma_{D_1}=18.9^{+4.6}_{-3.5}$ MeV,\cite{pdg}
even though a $1^+_{1/2}$ component can be contained in such physical
state due to the mixing allowed for the finite value of the charm quark
mass. Actually, another state with $J^P=1^+ $ has been 
identified$\;$\cite{anderson} with
mass $m_{D^{*0}_1}=(2461\pm^{41}_{34} \pm 10 \pm 32)$ MeV.
Both the states $2^+_{3/2}$ and $1^+_{3/2}$ decay to hadrons by
$d-$wave transitions, which explains their narrow width;
the strong coupling constant governing their two-body decays
can be determined using  experimental information.\cite{falk92}
On the other hand, 
the strong decays of the  states belonging to the $s_\ell^P={1
\over 2}^+$ doublet ($D_0,D_1^\prime$) 
occur through $s$-wave transitions,
with larger expected 
widths than in the case of the
doublet ${3\over 2}^+$. Indeed, analyses of
the coupling constant governing the two-body hadronic transitions
in QCD sum rules predict
$\Gamma(D_0^0 \to D^+ \pi^-) \simeq 180$~MeV and
$\Gamma(D_1^{\prime0} \to D^{*+} \pi^-) \simeq 165$~MeV.\cite{colangelo95}
Estimates$\;$\cite{colangelo95,kilian92} of the mixing
angle $\alpha$ between $D^*_1$ and  $D_1$ give $\alpha \simeq
16^0$.
Also in the case of $D_s$ and $B$
mesons experimental evidences of these states have been
reported.\cite{lep,ciulli}

Again we can introduce universal form factors, one for each transition
between couples of doublets. The universal function describing the
transition of an element of the fundamental doublet $(P,P^*)$ to the
doublet $(P_0,P_1^\prime)$ with $s_\ell=1/2$ will be referred to as
$\tau_{1/2}(v \cdot v^\prime)$, while the transitions of the fundamental
doublet to the $(P_1,P_2)$ one with $s_\ell=3/2$ will be described by the
function $\tau_{3/2}(v \cdot v^\prime)$. However, the heavy quark
symmetry does not predict the normalization of these functions, though
a great  simplification is again obtained, since the two $\tau$
functions take the place of 14 form factors, usually employed to describe
these transitions.

We can stress again what has been said in the beginning: HQET is
very powerful in providing us with  relations among different
quantities entering in the description of heavy hadron processes, and the
universal form factors are an important example. However, these quantities
cannot be computed within the same context, since they represent the
overlap of the light degrees of freedom. They are essentially 
nonperturbative objects, and should be computed by some other means. We shall
see that they will play a role in the subsequent analyses.

A very useful way to derive the structure of the matrix elements in HQET
is to combine the members of a given doublet in a single wave
function. \cite{casal}
For example, for the fundamental doublet one can write
\be
H_a={1+ \spur v \over 2}[P_{a \mu}^* \gamma^\mu-P_a \gamma_5]
\label{ha} \;\;,
\ee
\noindent where the first term represents the vector meson, the second the
pseudoscalar one, and $a$ is a light flavor index. The operators
$P_{a \mu}^*,~P_a$
destroy a vector and a pseudoscalar
meson, respectively, and contain a factor $\sqrt{m_P}$. Besides, $v^\mu
 P_{a \mu}^*=0$. Explicitly one can write
\bea
P(v) &=& -\sqrt{m_P}{1+ \spur v \over 2}\gamma_5  \; , \nonumber \\
P(v, \epsilon) &=& \sqrt{m_P}{1+ \spur v \over 2}\spur \epsilon 
\;\;.
\eea
Finally, ${\bar H}_a=\gamma_0 H_a^\dag \gamma_0$.
Let us now consider the transition between two heavy mesons $P(v) \to
P^\prime(v^\prime)$ induced by an external current coupled to the heavy
quarks, $J=h^{Q^\prime}_{v^\prime} \Gamma h^Q_{v}$, where $\Gamma$
is a product of Dirac matrices appropriate to the process at hand. Taking
into account the previous considerations on the factorization of the
overlap of the light degrees of freedom, it is possible to derive the most
general decomposition for the transition matrix element, satisfying all
the symmetry requirements (Lorentz
covariance, heavy quark symmetry and parity),
\be
\langle P^\prime(v^\prime)|J|P(v)\rangle =Tr[\Xi(v,v^\prime){\bar
P}^\prime(v^\prime) \Gamma P(v)]
\;\;, \label{trace}
\ee
\noindent where $\Xi(v,v^\prime)$ is a matrix containing the overlap of
the light degrees of freedom as well as the right tensorial structure 
to contract  free indices in the meson wave functions in order to
reproduce the correct Lorentz structure of the matrix element.
In the case of $ B \to D^{(*)}$ semileptonic transitions, one has
$\Xi(v,v^\prime)=-\xi(v \cdot v^\prime)$, and the following matrix
elements can be worked out,
\be
\langle D(v^\prime)| {\bar h_{v^{\prime}}}^c \gamma^{\mu} h_v^b 
|{\bar B} (v)\rangle  =
 \xi(v \cdot v^{\prime}) \;\sqrt{m_D m_B} \;( v^{\mu} + v^{\prime \mu})  
\;\;,  \label{xi1} 
\ee
and
\bea
&&\langle D^*(\epsilon^{\prime}, v^{\prime})| {\bar h_{v^{\prime}}}^c
\gamma^{\mu} (1- \gamma_5) h_v^b|
 {\bar B}(v)\rangle  \label{xi2} \nonumber\\[0.2cm]
&=& i \xi(v \cdot v^{\prime}) \;\sqrt{m_D m_B}\nonumber\\[0.2cm]
&\times& \large[ \epsilon^{\nu \alpha \mu \beta} 
\epsilon^{\prime *}_{\nu} v^\prime_\alpha v_\beta  
+ (1 + v \cdot v^{\prime})~ 
\epsilon^{\prime * \mu} - 
\epsilon^{\prime * \nu} v_\nu v^{\prime \mu} \large]
.
\eea 
The two positive parity doublets introduced before can be
represented
analogously to Eq.~(\ref{ha}). The $s_\ell=1/2$ doublet is described by the
following matrix:
\be 
S_a={1 + {\spur v} \over 2} (P^{\prime \mu}_1 \gamma_{\mu} 
\gamma_5 - P_0) \;\;, \label{sa} 
\ee
\noindent while the $s_\ell=3/2$ one is given by
\be 
T_a^{\mu} = {1 + {\spur v} \over 2} \Bigg\{ D_2^{\mu \nu} \gamma_{\nu}
- 
\sqrt{{3 \over 2}} D_1^{\nu} \gamma_5 \left[ g^{\mu}_{\nu} - {1 \over 3} 
\gamma_{\nu} (\gamma^{\mu} - v^{\mu})\right] \Bigg\} \;\;. 
\label{ta} 
\ee
\noindent The constructions could be further extended to higher
spins.\cite{falk0}
Let us now demonstrate in detail the normalization of the Isgur-Wise
function, since it will play an important role in the sequel. 
As it can easily be checked, for equal velocities, the vector current ${\bar
h}_v \gamma_\mu h_v$ in the
effective theory is conserved. Therefore, on account of Noether's
theorem, there is a
conserved charge, $N=\int d^3x h^\dagger_v(x) h_v(x)$. 
At zero recoil, using  invariance under translations, one has
\be
\int d^3x \langle P(v)|{\bar h}_v^\dagger(x) h_v(x)|P(v)\rangle  =
\int d^3x\langle P(v)|{\bar h}_v^\dagger(0) h_v(0)|P(v)\rangle  \nonumber \;\;. 
\ee
\noindent The operator ${\bar h}_v^\dagger h_v$ is the ``number" operator,
which counts the number of heavy quarks. Therefore, we have
\be
\int d^3x\langle P(v)|{\bar h}_v^\dagger(0) h_v(0)|P(v)\rangle  =V ~ \langle P(v)|P(v)\rangle =V~2
M_P v^0 \;\;,
\label{uno} 
\ee
\noindent  where $V$ is a normalization volume. On the other hand, using
Eq.~(\ref{xi1}), it holds also
\be
\int d^3 x \langle P(v)|{\bar h}_v^\dagger(x) h_v(x)|P(v)\rangle =
V~ \xi(1) ~2 M_P v^0 \label{due} \;\; .
\ee
\noindent Comparing Eqs.~(\ref{uno}) -- (\ref{due}), it follows that 
\be
\xi(1)=1 \label{normxi}\;.
\ee
Symmetry arguments have been used in physics more than once in order to
derive interesting phenomenological quantities. For example, the
 value of the vector coupling $g_V=1$ in the nuclear beta decay of
the neutron, which follows from assuming exact strong isospin
invariance of nuclear interactions, has allowed to  precisely extract
$V_{ud}$ from such a process. Moreover, 
in the limit of exact $SU(3)_F$ symmetry,
the form factor describing the
semileptonic decay $K \to \pi \ell \nu_\ell$, 
is normalized at zero recoil, receiving corrections only at the 
the second order in the symmetry breaking parameter $m_s-m_u$.
\cite{ademollo} This result, together
with the analysis of such ${\cal O}(m_s-m_u)^2$ corrections,\cite{roos} 
has given the possibility to obtain an accurate determination of
$V_{us}$. We shall see that the normalization of the Isgur-Wise function
will play an analogous role in the extraction of $V_{cb}$ from
experimental data.

\subsection{Matching}
All our previous considerations have been derived in the heavy quark
limit, $m_Q \to \infty$. For energies much below $m_Q$ we could assume
that HQET represents a good approximation to reality, in the same sense
that Fermi theory does for energies much below $M_W$. However, what should
be supplied in both cases are short distance corrections, which, as 
already noticed, provide new operators and modify the 
Wilson coefficients.
In this way we reach an important goal: we can exploit HQET in the long
distance regime, where we do not know how to cope with QCD; on the other
hand, the corrections to the  asymptotic predictions can be
computed sistematically in perturbative QCD. 

It can be easily understood why the ultraviolet behaviour of the two
theories is different: the typical terms arising in short distance
computations involve the logarithms of the heavy quark mass, exactly as
it was for the $W$, and therefore these logs diverge when taking the $m_Q
\to \infty$ limit. This is particularly important when one considers 
those operators, such as the vector or the axial-vector current, which 
 do not require  renormalization in QCD. In fact, these
operators do require renormalization in HQET.

Let us consider one such operator, $J_{QCD}$. We have seen that  it can be
expressed in terms of operators in the effective theory, which arrange
themselves in an expansion in the inverse $m_Q$ powers,
\be
J_{QCD}=J^{(0)}_{HQET}+{1 \over m_Q} J^{(1)}_{HQET}+\dots
\label{expj} \;\; ;
\ee
\noindent this equality should be understood in the sense of matrix
elements, since, according to the  discussion in Sec.~\ref{hqt}, the
states on which they act are different, those appearing in the matrix
element of the operator on the left-hand
side being dependent on $m_Q$, on the contrary of those on which the
operators on the right act.
Each term of the
expansion receives further short distance corrections. As a consequence,
at each order in ${1 / m_Q}$, several operators 
contribute to Eq.~(\ref{expj}), multiplied by suitable Wilson coefficients,
 so that each
$J^{(i)}$ in such equation will become $J^{(i)}=\sum_j
C_j(\mu)~J^{(i)}_j(\mu)$.
Let us focus our attention just on the first term in Eq.~(\ref{expj}), and
drop for simplicity the superscript $0$ and the subscript $HQET$, where
from now on the operators without indices will refer to HQET. 
Equation~(\ref{expj}) becomes
\be
J_{QCD}=\sum_i C_i(\mu)~J_i(\mu)+{\cal O}\left({1 \over m_Q}\right)=
\sum_i C_i(\mu)~Z_{ij}^{-1}(\mu)
~J_j^{bare}+{\cal O}\left({1 \over m_Q}\right)
\label{sd} \;\;,
\ee
\noindent where we took into account the fact that HQET currents require
renormalization by introducing a matrix of renormalization constants
$Z_{ij}$. This is a matrix because in general we could have different
operators with the same quantum numbers contributing at each order, which
could mix under renormalization. 
Since both $J_{QCD}$ and $J^{bare}$ are $\mu$-independent quantities, the
scale dependence should cancel between $C_i(\mu)$ and $Z_{ij}(\mu)$.
This gives rise to the renormalization group equation for the Wilson
coefficients
\be
\mu \left({d \over d \mu}- \hat \gamma^T \right) C(\mu)=0
\label{rge} \;\;,
\ee
\noindent where the coefficients have been collected in a single vector
$C$, and $\hat \gamma^T$ is the transpose of the matrix of anomalous
dimensions, obtained by the matrix ${\hat Z}$ of the renormalization
constants,
\be
\hat \gamma={\hat Z}^{-1} \mu {d \over d \mu}{\hat Z}
\label{anomdim} \;\;.
\ee
\noindent At a suitable high scale  $m\simeq {\cal O}(m_Q)$, one can
obtain the
Wilson coefficients by comparing the diagrams including gluon radiative
corrections in the
full and in the effective theory at the desired order in $\alpha_s$,
since in this case there are no more large logs and usual perturbation
theory works well. 
In this way, one can write the solution of Eq.~(\ref{rge}) as follows: 
\be
C(\mu)=U(\mu, m)C(m)
\label{solrge} \;\;.
\ee
\noindent Let us consider this solution in the case without
mixing, so that $\gamma$ is a number. This is a relevant example because,
as we shall see later, this is just the case of the vector and axial
vector currents, where no mixing occurs, and hence the anomalous dimension
is a diagonal matrix, allowing us
to apply the considerations which follow.
In this situation, one obtains
\be
U(\mu,m)= \left[ {\alpha_s(m) \over \alpha_s(\mu)} \right]^{\gamma_0 \over
2 \beta_0} \left[ 1+ {\alpha_s(m)-\alpha_s(\mu) \over 4 \pi} S + \dots 
\right] \label{u} \;\;,
\ee
\noindent where 
\be
S={ \gamma_1~ \beta_0 - \beta_1~\gamma_0 \over 2 \beta_0^2}
\label{s} \;\;,
\ee 
\noindent and 
$\gamma_0,~\gamma_1$ and $\beta_0,\beta_1$ are the first two coefficients
of the perturbative expansion of the 
anomalous dimension $\gamma$ and of the QCD $\beta$-function.
The approximation of considering only the first factor in
Eq.~(\ref{u}) corresponds to the leading log approximation, in
which the large logs have been resummed. Introducing the
one loop expression of $\alpha_s$ and expanding again this term, one
recovers the  perturbation theory result. The second term 
is the next-to-leading log correction, which requires the knowledge of
the second coefficient of the anomalous dimension.

If $C(m)$ has been computed at ${\cal O}(\alpha_s)$,
$$ 
{C(m)=C_0+C_1{\alpha_s \over 4 \pi}}\,,
$$ 
it is possible to
write the solution in
Eq.~(\ref{solrge}) separating the $\mu$ dependence and the heavy mass
dependence,
\be
C(\mu)=\hat C(m) \hat K(\mu) \;\;, \label{cmu}
\ee
\noindent where
\bea
\hat C(m) &=& \big[\alpha_s(m)\big]^{\gamma_0/2 \beta_0}
\left[C_0+(C_0~S+C_1){\alpha_s(m) \over 4 \pi} \right]  \; , \nonumber \\
\hat K(\mu) &=& \big[\alpha_s(\mu)\big]^{-\gamma_0/2 \beta_0} 
\left( 1 -{\alpha_s(m) \over 4 \pi}~S \right) 
\label{sep} \;\;.
\eea
We shall see in the following that this formalism allows a precise study
of the heavy hadron transitions, with
relevant phenomenological consequences.

\section{Leptonic Decay Modes and Heavy Meson Leptonic Constants}
The leptonic decay constants are non perturbative parameters which
enter universally in the description of bound state effects in a given
particle. For a pseudoscalar and
a vector particle, respectively, they are defined by the matrix elements
\bea
\langle 0|A_\mu|P(p)\rangle &=& i f_P p_\mu \; ,  \nonumber \\
\langle 0|V_\mu|P^*(p,\lambda)\rangle &=& i f_{P^*} m_{P^*} \epsilon_\mu(p,\lambda)
\nonumber \;\;,
\eea
\noindent
where $\epsilon$ is the polarization vector of the $P^*$ meson.
The leptonic constant  $f_P$  allows the complete description of the
purely leptonic decay mode of the pseudoscalar meson, with the
decay rate given by:
\be
\Gamma(P \to \ell^- {\bar \nu}_\ell) = {G_F^2 \over 8 \pi} |V_{q_1 q_2}|^2
f_P^2  \left( {m_\ell \over m_P}\right)^2 m_P^3 \left( 1- {m_\ell^2 \over
m_P^2} \right)^2
\label{leprate} \;\;. 
\ee
Therefore, the leptonic decay modes are the best
places  where  to experimentally measure
such hadronic parameters; indeed,  in the case of pion
the 
constant
$f_\pi$ is precisely determined from the process 
$\pi^- \to \mu^- {\bar \nu}_\mu$. 
However, in the case of heavy mesons like the $B$ meson, 
things are  hard. 
Purely leptonic decay channels present the difficulty
of the helicity suppression factor
$${\left( {m_\ell \over m_B} \right)}^2\,,$$ that makes the
purely
leptonic decay mode hardly accessible in the electron and in the muon
case (branching fractions of order $10^{-10}$, $10^{-7}$, respectively). 
In the case of the $\tau$ lepton, the helicity suppression is
absent, but the experimental $\tau$ reconstruction  is a difficult task. 

One could hardly overemphasize the importance of the
measurement of $f_B$,
due to the role played by this parameter in the description
of CP violation in the neutral $B$ system. 
Besides, as stems from Eq.~(\ref{leprate}), $f_B$
often appears together with some CKM matrix element, so that the
extraction
of either the decay constant or the relevant CKM element is affected by
the uncertainty in the other parameter. This is especially important in
the case of $V_{ub}$, which   is still affected by  quite a
large uncertainty.

Adopting arguments inspired by HQET, other
possibilities have been proposed, which  avoid 
helicity suppression.\cite{bmunugamma} The arguments are
based on the quark scaling behaviour of the pseudoscalar and the vector
$\bar q Q$ decay constants. As a matter of fact, 
using the trace formalism  as well as Eq.~(\ref{ha}), 
one can show, in a  straightforward manner,
that in the limit of large $b$ quark mass
\be
f_B=f_{B^*}={F \over \sqrt{m_B}} \;\;. \label{fhat}
\ee
The constant $F$ is another example of a universal
parameter in HQET: in the heavy quark limit, it describes both
pseudoscalar and vector heavy meson decays, irrespectively of their
flavor.  The simple relation in Eq.~(\ref{fhat}) receives both short
distance and ${\cal O}(m_Q^{-1})$ corrections; in the leading log
approximation, one finds\cite{hqet_1,politzer,neub1076,ji}
\be
{f_B \over f_D}=\sqrt{m_D \over m_B} \left( { \alpha_s(m_c) \over
\alpha_s(m_b) }\right)^{6/25} 
\ee
and
\be
{f_{P^*} \over f_P}= 1- {2 \alpha_s(m_Q) \over 3 \pi} \;\;,
\ee
with $P=D,~B$. This observation inspired  the proposal$\;$\cite{bmunugamma}
of using radiative leptonic $B$ decays $B \to \ell \nu \gamma$, in addition 
to purely leptonic modes, to get information on $F$ and $f_B$.
As a matter of fact,  $B \to \ell \nu \gamma$ processes are not helicity
suppressed, and their branching ratio, in the case of electron and muons
in the final state, is estimated to be one order of magnitude larger than
$\Gamma(B \to \mu \nu)$. A determination of $F$ by this mode, although
affected by some amount of theoretical error, would indeed be very precious
for the whole $B$ meson phenomenology.

\section{Isgur-Wise Function $\xi(y)$ and the Determination of $|V_{cb}|$}
Let us now consider the exclusive decays $B \to D^{(*)} \ell \nu$.
The weak matrix elements relevant to these transitions can be conveniently
parametrized in terms of form factors,
\bea
{\langle D(v^\prime)|V_\mu| B(v)\rangle  \over \sqrt{m_B m_D} }
&=&~h_+(y)~(v+v^\prime)_\mu+
h_-(y)~(v-v^\prime)_\mu  \; , \nonumber\\[0.2cm]
{\langle D^*(v^\prime,\epsilon)|V_\mu| B(v)\rangle \over \sqrt{m_B m_{D^*}} }
&=&i~h_V(y)~ \epsilon_{\mu \nu \alpha
\beta}\epsilon^{*\nu} v^{\prime \alpha} v^\beta \label{hi}  \; , \\[0.2cm]
{\langle D^*(v^\prime,\epsilon)|A_\mu| B(v)\rangle \over \sqrt{m_B m_{D^*}} }
&=&h_{A_1}(y)(y+1)\epsilon^*_\mu-
\big[h_{A_2}(y) v_\mu+h_{A_3}(y)v^\prime_\mu \big]\epsilon^* \cdot v \; , 
\nonumber 
\eea
where $v$ and $v^\prime$ are the initial and final meson 
four-velocities, respectively, and
$y=v \cdot v^\prime$. These are the form factors that,
in the $m_Q \to \infty$ limit, can all be expressed in
terms of a
universal function $\xi(y)$ introduced in Sec. 2. 
Comparing Eq.~(\ref{hi}) with
Eqs.~(\ref{xi1}) -- (\ref{xi2}), one finds
\bea
h_+(y)&=&h_V(y) = h_{A_1}(y) = h_{A_3}(y) = \xi(y)  \; , \nonumber \\[0.2cm]
h_-(y)&=&h_{A_2}(y) = 0 \label{relxi} \;\;.
\eea
Considering short distance corrections, 
all the form factors are related to a single function
in the  following, more complicated way:
\bea
h_+(y) &=& \left\{  C_1(y,\mu)+{y+1 \over 2} \left[ C_2(y,\mu)+
C_3(y,\mu) \right] \right\}~\xi(y,\mu)  \; , \nonumber \\
h_-(y) &=& {y+1 \over 2} \left[  C_2(y,\mu)-  C_3(y,\mu) \right]
 ~\xi(y,\mu)  \; , \nonumber \\
h_V(y) &=&  C_1(y,\mu)~\xi(y,\mu)  \; , \nonumber \\
h_{A_1}(y) &=&  C_1^5(y,\mu)~\xi(y,\mu) \; ,  \nonumber \\
h_{A_2}(y) &=& C_2^5(y,\mu)~\xi(y,\mu)  \; , \nonumber \\
h_{A_3}(y) &=& \left[  C_1^5(y,\mu)+
C_3^5(y,\mu) \right]~\xi(y,\mu) \label{cixi} \;\;.
\eea
\noindent In the previous equations, $ C_i^{(5)}(y,\mu)$ represent the
short
distance coefficients connecting the QCD vector (axial-vector) current to
the corresponding HQET current; there are three such coefficients because
now
we have three possible structures, i.e. $\gamma_\mu$,
$v_\mu,~v_\mu^\prime$. 
The $\mu$-dependence is the
same for all the functions $C_i^{(5)}$; 
therefore, according to Eq.~(\ref{cmu}),
one can extract such a dependence
by writing $\;$\cite{neubpr}
\be
C_i^{(5)}(y,\mu)={\hat C}_i^{(5)}(m_b,m_c,y) \; K_{hh}(y,\mu)
\label{nextcoef}
\ee
where $$K_{hh}=[\alpha_s(\mu)]^{-a_{hh}(y)} \Big\{1-{\alpha_s(\mu) \over
\pi}
Z_{hh}(y) \Big\}\,,$$ 
with $a_{hh}={2 \over 9} \gamma(y)$,
and 
$\gamma(y)$ is related to the
velocity-dependent anomalous dimension of the heavy-heavy current in
HQET\cite{falk90}
\bea
\gamma(y)&=&{4 \over 3} [y r(y)-1]  \; , \nonumber \\
r(y)&=& { \ln \left(y +\sqrt{y^2-1} \right) \over \sqrt{y^2-1} }
\label{gammadiy} \;\;.
\eea
As for the coefficient $Z_{hh}$, we refer the reader to the original
literature where it has been derived.
\cite{neubpr,neubert92,ji,kor} 
The expressions of the coefficient functions
${\hat C}_i^{(5)}$ are known at 
the next-to-leading order;\cite{neub92,neubpr}
it the leading-log approximation they simply read
$${\hat C}_1={\hat C}_1^5=\Big[{\alpha_s(m_b)\over
\alpha_s(m_c)}\Big]^{a_I}
[\alpha_s(m_c)]^{a_{hh}}\,,$$
with $a_I=-{6\over 25}$, the coefficients  $\hat C_2^{(5)}$ and
$\hat C_3^{(5)}$ being zero.

The $\mu$-dependence of the coefficients $C^{(5)}_i$ should cancel against
the function $\xi(y,\mu)$, so that the form factor
\be
\xi_{ren}(y)=K_{hh}(y,\mu)\xi(y,\mu) \label{xiren}
\ee
is  a renormalization group invariant function.
Notice that the pattern of radiative corrections do not
modify the normalization $\xi(1)=1$. 
To quantify this statement, let us introduce two combinations of
short distance coefficients which will be relevant in the following
discussion,
\be
\eta_V=\sum_1^3~C_i(y=1), \hskip 1 cm \eta_A=C_1^5(y=1)
\label{etas} \;\;.
\ee
\noindent For the $b \to c$ transition, the explicit
calculation$\;$\cite{neub92} gives
\be
\eta_V=1.025 \pm 0.006, \quad  \eta_A=0.986 \pm 0.006
\label{risetas} \;.
\ee
\noindent However, in the case of flavor conserving heavy quark currents,
one has 
\be
\eta_V=1,
\quad  \eta_A=1-{2 \alpha_s(m_Q) \over 3 \pi}\,,
\ee
which explicitly shows that the flavor conserving vector
current is not renormalized, assuring the $\xi$ normalization at zero
recoil point.

${{\cal O}\left( {1 /m_Q }\right)}$ corrections modify
this picture, introducing new operators in the expansion of the
QCD currents in terms of the HQET ones. However, the matrix elements of
these new operators all vanish in the zero recoil point at this order, so that
the normalization of the matrix elements in Eq.~(\ref{hi}) is further
preserved, receiving corrections only
at order $1 / m_Q^2 $. This is the content of 
Luke's theorem.\cite{luke}
Such  a theorem assures that the two  form factors $h_+,~h_{A_1}$in
Eq.~(\ref{hi}) receive only second order corrections in the inverse heavy
quark mass expansion at zero recoil point,
\bea
h_V(1)&=&\eta_V+{\cal O}\left( {1 \over m_Q^2 }\right)  \; , \nonumber \\
h_{A_1}(1)&=&\eta_A+{\cal O}\left( {1 \over m_Q^2 }\right) \label{luke}
\;\;.
\eea
\noindent
However, this does not hold for all the form factors appearing in
Eq.~(\ref{hi}), which simply do not contribute at $v=v^\prime$ because
they are multiplied by vanishing kinematical factors.\cite{rieckert}

This analysis is particularly relevant since the exclusive decays $B \to
D^{(*)} \ell \nu$ allow to access the CKM matrix element
$V_{cb}$. However, due to the
explained pattern of corrections, only the
decay to the charmed vector meson is suitable, since in this case
the rate, near the zero recoil point, is proportional to $|h_{A_1}(y)|^2$,
which is protected by  Luke's theorem. This is not the case for the
transition to the $D$ meson.\cite{neub91}

The differential decay rate for the process $B \to D^* \ell \nu$ can be
written as follows:
\bea
{d \Gamma \over dy} (B \to D^* \ell \nu)&=& {G_F^2 \over 48 \pi^3}
(m_B-m_{D^*})^2 m_{D^*}^3 \sqrt{y^2-1} (y+1)^2 \label{spec}\\[0.2cm]
&\times&
\left[ 1+{4y \over y+1} {m_B^2-2y m_B m_{D^*}+m_{D^*}^2 \over
(m_B-m_{D^*})^2} \right] |V_{cb}|^2 {\cal F}^2(y) 
\;\;, \nonumber
\eea
\noindent where the function ${\cal F}^2(y)$ differs from the Isgur-Wise
function for short distance and ${\cal O} (m_Q^{-1})$ corrections. What
should be noticed is that: {\it i}) the normalization of $\cal F$ at $y=1$
is known; {\it ii}) the spectrum vanishes at $y=1$.
The comparison of the experimental data with the previous result could be
translated in a result for the product $|V_{cb}|^2 {\cal F}(1)$, 
after a suitable extrapolation to the zero recoil point due to the
vanishing of the spectrum. This has lead to an accurate determination of
$|V_{cb}|$, which can be considered among the most important
applications
of the heavy quark symmetries. We quote the last
result,\cite{vcbwg}
\be
|V_{cb}|=(39.8\pm 1.8 \pm 2.2) \times 10^{-3} 
\label{vcb} \;\;.
\ee
\noindent

\section{Determination of the Universal Form Factors from the QCD Sum Rules}
 HQET allows one to derive model independent relations,
holding in the heavy quark limit, mainly on the basis of the decoupling of
the light degrees of freedom from the spin of the 
heavy quark. We have noticed that in
such a limit it is possible to introduce a
whole series of nonperturbative parameters with a universal meaning,
such as the decay constant $ F$, or the universal form factors
describing weak decays among the members of doublets of asymptotically
degenerate states. The usefulness of such relations is
evident in the case of the Isgur-Wise function, where one can use symmetry
arguments in order to predict the zero recoil point normalization and
can use
this information to extract $V_{cb}$ from experimental data. 
However, we often need something more, i.e. the quantitative estimate of
these universal quantities, which cannot be done within HQET, but requires
the use of some nonperturbative technique, such as QCD sum rules or
lattice QCD. In particular, the use of QCD sum rules in the framework of
HQET revealed a very powerful tool and many HQET parameters have been
estimated in this context. We shall review in the following subsections
the determination of the Isgur-Wise function
$\xi$$\;$\cite{rad,neubert92,neub93}
and the  computation of the analogous universal form factors
$\tau_{1/2},~\tau_{3/2}$, describing the
transitions of a $B$ meson to excited positive parity charmed
mesons.\cite{lorotau}$^{\!,\,}$\cite{noitau}
As we shall see, a preliminary part of such computations is the
determination, within the same QCD sum rule framework, of the leptonic
constants of the heavy mesons. The functions $\xi$ and $\tau_{1/2}$ have
been computed  including ${\cal O}(\alpha_s)$ corrections, so that we
shall have the chance to discuss the role of such
corrections in practical situations.

Let us summarize the basic derivation of a generic universal form factor
using QCD sum rules.
The starting point is a three-point correlator, defined in the effective
theory,
\begin{eqnarray}
\Pi_{1,2}(\omega, \omega^\prime, y) &=& i^2 \int dx \; dz ~e^{i(k' x-k z)}
\langle 0|T[ J_1^{v^\prime}(x) J_W(0) J_2^{v}(z)^\dagger]|0\rangle 
\nonumber \\[0.1cm]
&=&  \Pi(\omega, \omega^\prime, y)~ Tr \left\{\Gamma_1~ {1 + {\spur 
v^\prime} \over 2} \Gamma~ {1 +{\spur v} \over 2}
\Gamma_2^\dagger \right\}
\label{threep}
\end{eqnarray}
where $J_1^{v^\prime}={\bar q} \Gamma_1
h^{Q^\prime}_{v^\prime}$, 
$J_2^v={\bar q} \Gamma_2 h^Q_v$ are the  effective currents
interpolating the heavy mesons with heavy quarks $Q^\prime$ and $Q$
respectively.\footnote{The issue of the choice of the interpolating
currents in HQET is extensively discussed by Dai {\it et al}.\cite{dai}}
Moreover,  $J_W=h^{Q^\prime}_{v^\prime}\Gamma h^Q_v$ is the weak
current corresponding to the transition 
$h^Q_v \to h^{Q^\prime}_{v^\prime}$. The variables $k, k^\prime$ are the 
residual momenta, obtained by
the expansion of the heavy meson momenta in terms of the four-velocities:   
$P=m_Q v+ k$, $P^\prime=m_{Q^\prime} v^\prime+ k^\prime$.

Using the analyticity of $\Pi(\omega, \omega^\prime, y)$ in the variables
$\omega=2 v\cdot k$ and $\omega^\prime=2 v^\prime\cdot k^\prime$ at fixed
$y$,
one can represent the correlator
(\ref{threep}) by a double dispersion
relation of the form
\begin{equation}
\Pi(\omega, \omega^\prime, y) = \int d \nu d \nu^\prime  
{\rho(\nu, \nu^\prime, y) \over (\nu - \omega - i \epsilon)
(\nu^\prime - \omega^\prime - i \epsilon) } \;\; \;, \label{correlatore}
\end{equation}
apart from possible subtraction terms. The correlator
$\Pi(\omega, \omega^\prime, y)$
receives contributions from poles located at positive real
values of $\omega$ and $\omega^\prime$, corresponding to the physical
single particle hadronic states
in the spectral function $\rho(\nu, \nu^\prime, y)$.

This contribution is proportional to the universal function
$\Xi$ appropriate to the 
transition at hand, through the following relation:
\begin{equation}
\Pi_{pole}(\omega, \omega^\prime, y) \propto
{ \Xi(y, \mu)  F_1(\mu)  F_2(\mu)  \over
(2 \bar \Lambda_1  - \omega - i \epsilon)
(2 \bar \Lambda_2 - \omega^\prime - i \epsilon) } \;, \label{pole}
\end{equation}
where $\mu$ is the renormalization scale
and  $F_1(\mu)$, $F_2(\mu)$ are the effective leptonic
constants of the heavy mesons interpolated by the currents
$J_1^{v^\prime}$ and $J_2^v$ respectively, in analogy to Eq.~(\ref{fhat}).
The mass parameters $\bar \Lambda_1$ and $\bar \Lambda_2$ identify the
position
of the poles in $\omega$ and $\omega^\prime$,
and can be interpreted as binding energies of the heavy mesons, in
accordance with Eq.~(\ref{hadmass}).

\begin{figure}[htb]
\hskip 0.5in
\psfig{figure=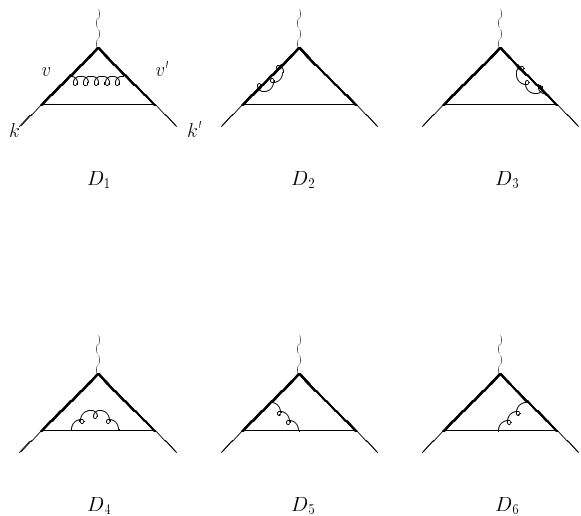,width=8cm}
\vspace{-2.7cm}
\caption{Two-loop diagrams relevant for the calculation of ${\cal O}
(\alpha_s)$
corrections to the perturbative part of the QCD sum rule for the universal
form factors. The heavy lines represent the heavy quark
propagators in HQET.
\label{diagrams}}
\end{figure}

The higher state contributions to $\rho(\nu, \nu^\prime,y)$ can be
taken into account by a QCD continuum starting at some thresholds $\nu_c$ and 
$\nu^\prime_c$,  and  are modeled by the perturbative spectral 
function $\rho^{pert}(\nu, \nu^\prime,y)$ according to the quark-hadron duality 
assumption.\cite{misha} 
Here, $\rho^{pert}$ is the absorptive part of the perturbative 
quark-triangle diagrams, with two heavy quark lines corresponding to the weak 
$Q \to Q^\prime$ vertex and one light quark line connecting the heavy
meson 
interpolating current vertices in Eq.~(\ref{threep}). At the
next-to-leading 
order in $\alpha_s$, all possible internal gluon lines in such triangle 
diagrams must be considered, as displayed in Fig.~\ref{diagrams}.

Therefore, for the dispersive representation in Eq.~(\ref{correlatore}) 
in terms of hadronic 
intermediate states one assumes the ansatz
\be
\Pi(\omega, \omega^\prime, y)= \Pi_{pole}(\omega, \omega^\prime, y) +
\Pi_{continuum}(\omega, \omega^\prime, y) \label{ansatz}
\ee
where, for simplicity,
the dependence of the continuum contribution on the thresholds
$\nu_c$ and $\nu^\prime_c$ has been omitted.

The correlator
$\Pi(\omega, \omega^\prime, y)$ can be expressed in QCD in the 
Euclidean region, 
i.e. for large  negative values of $\omega$ and $\omega^\prime$, in terms of 
perturbative and nonperturbative contributions
\begin{equation}
\Pi(\omega, \omega^\prime, y) = \int d \nu d \nu^\prime
{\rho^{pert}(\nu, \nu^\prime, y) \over (\nu - \omega - i \epsilon) 
(\nu^\prime - \omega^\prime - i \epsilon) }  + 
\Pi^{np}(\omega, \omega^\prime, y)
\;\;\;. \label{qcd}
\end{equation}
In Eq.~(\ref{qcd}),
$\Pi^{np}$ represents the series of power corrections in  the ``small"
${1 / \omega}$  and  ${1 / \omega^\prime}$ variables,
determined by quark and gluon vacuum condensates ordered by increasing 
dimension. These universal QCD parameters account for general properties of 
the nonperturbative strong interactions, for which asymptotic freedom
cannot be applied. The lowest dimensional ones can be obtained from
independent theoretical sources, or fitted from other applications of QCD sum 
rules in the cases where the hadronic dispersive contribution is particularly 
well-known.\cite{col} In practice, since the higher dimensional condensates are
not known, one 
truncates the power series and {\em a posteriori} 
verifies the validity of such an  approximation. 

The QCD sum rule 
is finally obtained by imposing that the two representations 
of $\Pi(\omega, \omega^\prime, y)$, namely the QCD representation
(Eq.~(\ref{qcd})) 
and the pole-plus-continuum ansatz (Eq.~(\ref{ansatz})), match in a suitable
range of 
Euclidean values of $\omega$ and $\omega^\prime$. 

A double Borel transform  in the variables 
$\omega$ and $\omega^\prime$
\begin{equation}
{1 \over \tau} \hat{\cal B}_\tau^{(\omega)}=  \lim \; 
{\omega^n \over (n-1)!} \big( - {d \over d \omega} \big)^n , 
\hskip 1cm (n\to \infty, \omega \to -\infty, \tau=- {\omega \over n} \; 
{\rm fixed} )
\end{equation}
(and similar for $\hat{\cal B}_{\tau^\prime}$)
is applied to ``optimize" the sum rule.
As a matter of fact, this operation has two effects. The first one consists of
factorially improving the convergence 
of the nonperturbative series, justifying the truncation procedure;
the second effect  enhances the role of the lowest-lying meson states 
while minimizing 
that of the model for the hadron continuum.  The 
{\it a priori} undetermined mass parameters $\tau$ and $\tau^\prime$ must
be chosen in a suitable range of values, 
expected to be 
of the order of the typical hadronic 
mass scale $(\ge 1\; GeV)$, where the optimization is verified and,
in addition, the prediction turns out to be 
reasonably stable.
After the Borel transformation, possible subtraction terms are eliminated and
 Eq.~(\ref{correlatore}) can be rewritten as
\be
\hat \Pi (\tau, \tau^\prime, y)=\int d \nu d \nu^\prime \; e^{-\Big({\nu \over 
\tau}+{\nu^\prime \over \tau^\prime}\Big)} \rho(\nu, \nu^\prime, y)\;. 
\label{bor}
\ee
\noindent Equation~(\ref{pole}) shows that the preliminary
evaluation of the constants $F_1(\mu)$ and $F_2(\mu)$ is necessary to
exploit  the sum rule for the determination of $\xi$. 
This is done within the same HQET QCD sum rule framework,
specifically by evaluating two-point functions. 

In the following we shall
give the results for these quantities focusing on the transitions $B
\to (D,D^*)$ and $B \to (D_0,D_1^\prime)$.

\subsection{The Isgur-Wise Function}
The main properties of the Isgur-Wise function have been already
described in the previous sections. Here we  recall the derivation of
such form factor using QCD sum
rules.\cite{rad,neubert92,neub93} As stressed
above, a preliminary part of the analysis consists of the
effective theory  calculation 
 of the leptonic constant  of the heavy mesons
belonging to the fundamental doublet $(P,P^*)$, $P=D,~B$, since it enters
in the sum rule for the $\xi$ function through the relation in
Eq.~(\ref{pole}).
We have already met such a
parameter in Eq.~(\ref{fhat}); the constant $F$ in
that
equation is defined by
\be
\langle 0|J_5^v|P(v)\rangle =F(\mu) \label{fdef} \;\;,
\ee
\noindent where $J_5^v={\bar q}i \gamma_5 h_v$ interpolates the heavy
pseudoscalar meson.
One considers therefore the two-point correlator
\be
\psi (\omega) =i \int d^4x e^{i k \cdot x} \langle 0|T \{ J_5^{v
\dagger}(x) J_5^v(0) \}|0\rangle  \;\;.
\ee
\noindent One could have chosen equally to work with the vector
current and the $P^*$ meson, due to the degeneracy of the two members of
the doublet.

At the
next-to-leading order in renormalization group improved perturbation
theory it is possible to define a renormalized scheme-independent
quantity as follows:
\be
F_{ren}=[\alpha_s(\mu)]^{2/9} \left\{ 1- {\alpha_s (\mu) \over \pi}
[Z+ \delta] \right\} ~F(\mu)
\label{fren} \;\;,
\ee
\noindent where, in the  $\overline{MS}$ scheme, 
$$Z=3{153-19n_f \over (33-2n_f)^2}-{381-30n_f+28\pi^2 \over 36 (33-2 n_f)}$$ 
and $\delta=2/3$.

The sum rule for $F(\mu)$ allows to evaluate also the parameter $\bar
\Lambda$. This can be done
considering the logarithmic derivative
of the sum rule itself with respect to the Borel
parameter $\tau^\prime$. The results are~\cite{neubert92,matthias92}
\be
F_{ren}=0.40 \pm 0.06~{\rm GeV}^{3/2}\,,\qquad {\bar \Lambda}=0.57 \pm
0.07~{\rm GeV} \;.
\label{resmat}
\ee
\noindent The important result is that radiative corrections modify by
$\sim 30
\%$ the value of $F_{ren}$, a result which is not specific of QCD sum
rules, but reflects a general property of the considered two-point
correlator, and whose origin could be traced back to a Coulomb interaction
between quarks, since the largest enhancement comes from the contribution
of gluon exchange between the heavy and the light quark. The important
role of radiative corrections in the numerical evaluation of $F$ is
confirmed by other analyses.\cite{broad92,bagan92}

The QCD sum rule 
result for $F$ can be used as an input in the three-point
function in Eq.~(\ref{threep}) in order to compute the Isgur-Wise
function. In this case one can choose the two Dirac structures
$\Gamma_1,~\Gamma_2$  to interpolate the ground state heavy mesons
$(P,P^*)$. Therefore, the choices $\Gamma_{1,2}=i \gamma_5$ (corresponding
to the pseudoscalar meson $P$) or $\Gamma_{1,2}=\gamma_\mu- v_\mu$
(corresponding to the vector meson $P^*$) are equally acceptable, since
the
structure resulting from the computation of the trace will factor out in
all the terms contributing to the sum rule. Finally, in Eq.~(\ref{threep}),
we choose  
$J_W=h_{v^\prime}^{Q^\prime}\gamma_\mu(1-\gamma_5)h_v^Q$, in order to
describe the weak process.

In the lowest order in 
the perturbation theory, the sum rule fulfills the
important constraint of reproducing the zero-recoil point normalization of
the Isgur-Wise function,\cite{rad,neubert92} a result which continues
to hold after the inclusion of the $\alpha_s$ correction.\cite{neub93}  Hence,
we observe that the QCD sum rules formulated in the framework of the effective
theory automatically incorporate the correct normalization of the
function $\xi$. This observation stresses the suitability of employing
this technique in such a context. Furthermore, the inclusion of radiative
corrections to a three-point correlator of heavy quark currents has been
done for the first time in the HQET, where the task is simplified due to
the  simpler Feynman rules. The calculation of the loop integrals required
to evaluate the diagrams in Fig.~\ref{diagrams} can be done 
using integration by parts, in such a way as
to reduce complicated integrals
to simpler ones by recursive relations, and the method of differential
equations developed by Kotikov.\cite{kotikov} The inclusion of radiative
QCD corrections to the quark condensate had been done before the full
inclusion of such corrections in the perturbative
term.\cite{neubert92} Also it has been shown$\;$\cite{neub93} that the
calculation of the relevant loop integrals is
easier  when dealing directly with the Borel
transformed sum rule. In the case of the Isgur-Wise function, the symmetry
of the correlator allows to set equal  the two Borel parameters:
$\tau=\tau^\prime$, a simplification which is not allowed in the case of
the transitions of the $B$ meson to excited states, as we shall see in the
next subsection.

Let us observe that the currents employed in 
Eq.~(\ref{threep}) are those defined in the effective theory and therefore
are subject to renormalization. 
When computing any universal form factor, 
in Eq.~(\ref{threep}) there appear two heavy-light currents and a
heavy-heavy current; in the ${\overline {MS}}$ scheme the renormalization
constants of such currents are
\be
{\cal Z}_{hl}=1- {\alpha_s \over 2 \pi {\hat \epsilon}}\,, \qquad
{\cal Z}_{hh}=1+{\alpha_s \over 2 \pi {\hat \epsilon}} 
\gamma(y) \label{renconst} \;\,
\ee
where
\be
{1 \over {\hat \epsilon}}={1 \over \epsilon} +\gamma_E- \ln 4 \pi,
\ee
the parameter $\epsilon$ being defined in the $D$-dimensional
space-time by 
$$D=4+2 \epsilon,$$ 
reflecting the use of dimensional
regularization in the employed renormalization scheme.
 $\gamma(y)$ has been defined in the previous section.
The inclusion of $\alpha_s$ corrections to
the three-point correlator should reproduce the singularity structure of
these renormalization constants. As a matter of fact, this is 
explicitly verified both in the case of the $\xi$ function and for the
function $\tau_{1/2}$ to be considered in the next section.

We will not report here the sum rule for $\xi$, referring the reader to
the original literature.\cite{neub93} What is worth noticing is that the
radiative corrections turn out to be small, due to their cancelation in
the ratio of three- to two-point functions.
 For this reason  not only the normalization of $\xi$ is preserved, but
also the impact of QCD corrections is modest  for large recoil.

\begin{figure}[htb]
\hskip 0.5in
\psfig{figure=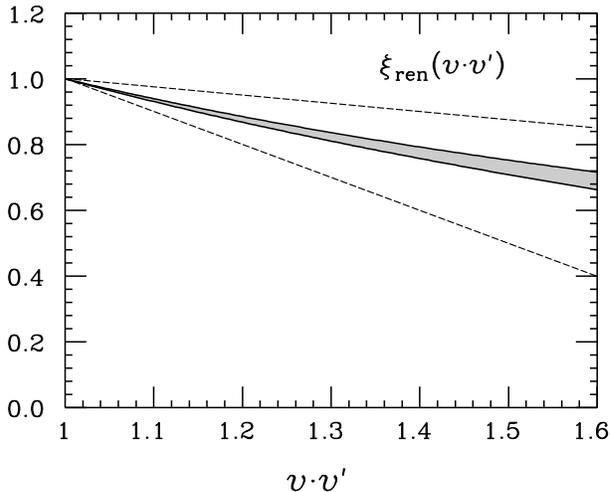,width=8cm}
\caption{QCD sum rule result for the Isgur-Wise function in the range of
$y=v \cdot v^\prime$ allowed in $B \to D^{(*)}$ transitions. The dashed
lines indicate the bounds on the slope at $y=1$ derived applying the
Bjorken and Voloshin sum rules. \label{xifig}}
\end{figure}

In Fig.~\ref{xifig} we show the QCD sum rule outcome\cite{neub93}
 for the renormalized
scheme-independent Isgur-Wise function.\footnote{The figure was taken  
from a review by M. Neubert.\cite{neubtasi}} 
It is possible to fit the curve by writing
\be
\xi(y)=\xi(1) \left[1-\rho^2~(y-1) \right]
\label{slope} \;\;,
\ee
\noindent where $\rho^2$ is known as the slope of the $\xi$ function. An
analogous decomposition holds for the function $\xi_{ren}$.
Hence, the sum rule predicts~\cite{neubpr}
\be
\rho^2_{ren}=0.66 \pm 0.05 \label{rhoren} \;.
\ee
This result is in agreement with the others available in
literature.\cite{blok,ball} In particular, as shown by
Neubert,\cite{neubtasi}
it lies within the bounds which
can be obtained on the basis of two sum rules, the Bjorken sum
rule$\;$\cite{bj90} and
the Voloshin sum rule.\cite{vol92,ural} Such two sum rules will
be discussed in the next section.

\subsection{$B$ decays to Excited Charmed Resonances}
It is worth analyzing other cases analogous to the determination of the
Isgur-Wise form factor  $\xi$, in particular it is interesting to
calculate the universal form factors governing the
semileptonic $B$ meson decays into the positive parity charmed excited
states. These higher-lying states correspond to the
$L=1$ orbital excitations in the non-relativistic constituent quark model.
Besides their theoretical relevance
to HQET,\cite{isgur91} in particular to the aspects of the QCD sum rule
calculations, such $B \to D^{**}$
semileptonic transitions (from now on $D^{**}$ will indicate the generic
$L=1$ charmed state)
have numerous additional points of physical interest.
Indeed, in principle these decay modes may
account for a sizeable fraction of semileptonic $B$-decays, and
consequently they represent a well-defined set of corrections to the
theoretical prediction that the total semileptonic $B\to X_c$ decay rate 
should be saturated by
the $B\to D$ and $B\to D^*$ modes,\cite{hqet_1} in the limit
$m_Q\to\infty$ and under the condition  $(m_b-m_c)/(m_b+m_c)\to 0$ (the
so-called small-velocity limit). 
Moreover, the shape of the inclusive differential decay distribution
in the lepton energy could reflect contributions from the
$B\to D^{**}$  modes.

The issue of the contribution of such modes to the $B$ semileptonic
branching ratio ${\cal B}_{SL}$ is also relevant since theoretical
determinations of this
quantity stay above the experimental data. In fact,   
the parton model gives ${\cal B}_{SL}\simeq 15 \%$, with negligible
${1 / m_Q}$ corrections.\footnote{As we shall see in the
next section, the
parton model result identifies with the leading term in the $1/m_Q$
expansion for inclusive decay rates.} 
Perturbative QCD corrections have a
more relevant impact, nevertheless leaving ${\cal B}_{SL} \ge 12.5 \%$. 
Also the experimental results seemed controversial for a long time, since
the data obtained at the $ \Upsilon (4S)$ by the CLEO Collaboration were
different from those obtained by the LEP Collaborations at the $Z^0$
peak. Recent results$\;$\cite{poling} show an improved situation,
\bea
{\cal B}(B \to X \ell \nu_\ell) &=& (10.63 \pm 0.17)\% \hskip 0.7 cm(Z^0
~~ {\rm average}) \; , 
\nonumber \\
{\cal B}(B \to X \ell \nu_\ell) &=& (10.45 \pm 0.21)\% \hskip 0.7 cm
(\Upsilon (4S) ~~ {\rm average}) \;\; ;
\eea
\noindent the last number is the Particle Data Group average for the
data at the $\Upsilon (4S)$.\cite{pdg}
A possible reduction of the theoretical prediction could be obtained
increasing the value of $n_c$, i.e. the number of charmed hadrons produced
in the $B$ decay. However, since the various experimental data for these
parameter show a very good agreement both among themselves and with
theoretical predictions, it is likely that the solution should be found
elsewhere. In particular, it is important to fully understand which is
the  contribution of the various possible decay modes, for example, those
with $D^{**}$ in the final state. 

Another important result, relevant both
to phenomenology and to the critical
tests of HQET, is the relation of the $B\to D^{**}$ form factors 
at zero recoil to the slope $\rho^2$ of the $B\to D^{(*)}$
Isgur-Wise function, through the Bjorken sum rule,\cite{bj90}
\be
\rho^2={1 \over 4}+\sum_n |\tau_{1/2}^{(n)}|^2+2 \sum_m
|\tau_{3/2}^{(m)}|^2 \label{bjorken}\;\;.
\ee
\noindent In this equation $n,~m$ identify the radial
excitations of states with the same $J^P$. The $D^{**}$ states considered
here represent the lowest lying states contributing to the left-hand side
of Eq.~(\ref{bjorken}).

Of similar interest for HQET is the test of the upper bound
on such universal form factors at zero recoil, involving the heavy meson 
``binding energy'' and the $D^{**}-D$ mass splittings,\cite{vol92,ural}
\be
{\bar \Lambda}=2 \left( \sum_n \epsilon_n |\tau_{1/2}^{(n)}|^2+2
\sum_m \epsilon_m |\tau_{3/2}^{(m)}|^2 \right)
\label{voloshin} \;\;,
\ee
\noindent where $\epsilon_k=M_{H_Q^{(k)}}-M_{P_Q}$.

Besides, the investigation of the
semileptonic $B$ transitions to excited charm states
is an important preliminary study for the theoretical analysis of the
production of such states in nonleptonic $B$
decays,\cite{nonlepb,anderson}
as well as for the identification of additional decay modes
(such as $D^{(*)} D^{(*)} \pi$) suitable for the investigation of CP
violating effects at $B$ factories.\cite{pene98}

Finally, as a byproduct of the QCD sum rule calculation, theoretical   
predictions for the  $D^{**}$ meson masses can be
obtained, which are obviously interesting {\it per se}.

We have already introduced  two universal form factors $\tau_{1/2}$,
$\tau_{3/2}$ describing the $B$ meson transitions to the members
of the two doublets with spin-parity $J^P_{s_\ell}=(0^+,1^+)_{1/2}$,
$J^P_{s_\ell}=(1^+,2^+)_{3/2}$ respectively. 
Let us consider now the HQET QCD sum rule 
calculation of the function $\tau_{1/2}$, which has been performed at
the
next-to-leading order in the
renormalization group improved perturbation
theory.\cite{noitau} The analogous result for the function
$\tau_{3/2}$ is 
available only at tree level in perturbative QCD
corrections.\cite{lorotau,dai,dai1}

The matrix elements of the semileptonic
$B \to D_0 \ell \bar \nu $ and $B \to D_1^\prime \ell \bar \nu$
transitions can be parameterized in terms of six form factors,
\bea
{\langle D_0(v^\prime)|{\bar c}\gamma_\mu \gamma_5 b|B(v)\rangle 
\over {\sqrt {m_B m_{D_0}}}  }
&=& g_+ (v+v^\prime)_\mu+
 g_- (v-v^\prime)_\mu  \; , \nonumber \\
{\langle D^\prime_1(v^\prime,\epsilon)|{\bar c}\gamma_\mu(1-\gamma_5)b|B(v)\rangle 
\over {\sqrt {m_B m_{D^*_1}}} }
&=& g_{V_1} \epsilon^*_\mu +\epsilon^* \cdot v \; [g_{V_2} v_\mu+
g_{V_3} v^\prime_\mu]  \; , \nonumber \\
&-&i \; g_A
\epsilon_{\mu \alpha\beta \gamma} \epsilon^{*\alpha} v^\beta v^{\prime
\gamma}
\;\;\;,  \label{full}
\eea
where $v$ and $v^\prime$ are four-velocities
and $\epsilon$ is the $D_1^\prime$ polarization vector.
The form factors $g_i$ depend on the variable $y=v \cdot v^\prime$, which
is directly related to the momentum transfer to the lepton pair.
Again, the heavy quark spin symmetry allows us
to relate the form factors
$g_i(y)$ in Eq.~(\ref{full}) to a single function 
$\tau_{1/2}(y)$~\cite{isgur91} through short-distance coefficients, 
perturbatively calculable, which depend on the heavy quark masses $m_b,m_c$,
on $y$ and on a mass-scale $\mu$. 
The heavy quark spin symmetry allows us
to  connect the QCD vector and axial-vector
currents to the HQET currents. At the next-to-leading
logarithmic approximation in $\alpha_s$ and in the infinite heavy quark
mass limit,
the relations between $g_i$ and $\tau_{1/2}$ are given by
\bea
g_+(y) + g_-(y) &=& -2 \;\Big( C_1^5(y,\mu)+(y-1) C_2^5(y,\mu)\Big)
\; \tau_{1/2}(y,\mu)  \; , \nonumber \\
g_+(y) - g_-(y) &=& 2\;\Big( C_1^5(y,\mu)-(y-1)C_3^5(y,\mu)\Big)
 \; \tau_{1/2}(y,\mu)  \; , \nonumber \\
g_{V_1}(y)&=&2(y-1) \;C_1(y,\mu) \; \tau_{1/2}(y,\mu) \; ,  \nonumber \\
g_{V_2}(y)&=&-2\; C_2(y,\mu)\; \tau_{1/2}(y,\mu) \; ,  \nonumber \\
g_{V_3}(y)&=&-2\;\Big(C_1(y,\mu)+C_3(y,\mu)\Big) \;\tau_{1/2}(y,\mu)  \; , \nonumber
\\
g_A(y)&=&-2 \; C_1^5(y,\mu) \; \tau_{1/2}(y,\mu) \;.
\label{rel_formf}
\eea
Following the same steps leading to Eq.~(\ref{xiren}), a
renormalization-group
invariant form factor can be defined
\be
\tau_{1/2}^{ren}(y)=K_{hh}(y,\mu) \tau_{1/2}(y,\mu)\;; \label{tauren}
\ee
hence, in Eq.~(\ref{rel_formf}) one can substitute $C_i^{(5)}(y,\mu)$ by
${\hat C}_i^{(5)}(y)$ and $\tau_{1/2}(y,\mu)$ by $\tau_{1/2}^{ren}(y)$.
Analogous relations hold for the eight form factors parameterizing the
matrix
elements of $B \to D_1 \ell \bar \nu$ and
$B \to D_2 \ell \bar \nu$; in this case the heavy quark symmetry allows
to
relate them  to the  universal function 
$\tau_{3/2}(y)$.\cite{isgur91} 
The main difference with respect to the Isgur-Wise form factor
$\xi(y)$ is that one cannot invoke symmetry arguments to predict the 
normalization of both $\tau_{1/2}(y)$ and $\tau_{3/2}(y)$, and therefore a
calculation of the form factors in the whole kinematical range is
required.
For $B \to (D_0, D^\prime_1) \ell \bar \nu$
the physical range for the variable $y$ is restricted between
$y=1$ and $y=1.309-1.326$, taking into account
the values for the mass of $D_0$, $D^\prime_1$
($m_{D_0,D^\prime_1}=2.40-2.45$ GeV).

Let us consider the transition $B \to D_0$ and define
\be
\langle 0| J_s^v |P_0(v) \rangle  = F^+(\mu) \label{fpiu} \;\;,
\ee
\noindent where $J_s^v={\bar q}h^{Q^\prime}_{v^\prime}$ represents the
local interpolating current of the scalar ($D_0$) meson.
In analogy to Eq.~(\ref{hadmass}), we can write, in the heavy quark limit,
\be
{\bar \Lambda}^+=M_{D_0}-m_c \label{lbarpiu} \;\;.
\ee
\noindent By considering the following two-point  correlator:
\be
\Psi(\omega^\prime)=i \int d^4x \; e^{ik^\prime\cdot x}
\langle 0|T[J_s^{v^\prime}(x) J_s^{v^\prime}(0)^\dagger]|0\rangle 
\;\;\;, \label{corr2pt}
\ee
\noindent it is possible to obtain a sum rule for the constant $F^+(\mu)$.
 A $\mu$-independent constant $F^+_{ren}$ can be
defined, using the relation
between $F^+(\mu)$ and the matrix element of the  scalar current in
full QCD,
\be
F^+_{ren}=[\alpha_s(\mu)]^{d \over 2} \Big[
1-{\alpha_s(\mu) \over \pi} \; Z] \;F^+(\mu)
\ee
where $Z$ has been defined in the previous subsection.

Again, the sum rule for $F^+(\mu)$ allows to evaluate the parameter
$\bar\Lambda^+$, by 
considering the logarithmic derivative with respect to the Borel
parameter $\tau^\prime$.
The following predictions are obtained:\cite{noitau}
\be
\bar \Lambda^+= 1.0 \pm 0.1 \; {\rm GeV} \,,\qquad
F^+_{ren}=0.7 \pm 0.2 \;{\rm GeV}^{3\over 2}\;. \label{fpiun}
\ee
\noindent Perturbative ${\cal O}(\alpha_s)$ corrections represent a
sizeable
contribution to the sum rule for $F^+_{ren}$, similar to the situation met
in the case of $F$.\cite{broad92,neubert92,bagan92}

\begin{figure}[htb]
\hskip 0.5in
\psfig{figure=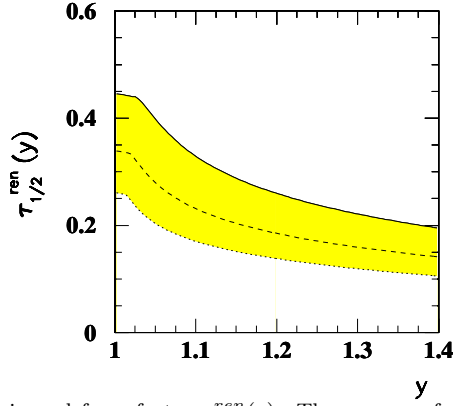,width=8cm}
\vspace{-2.cm}
\caption{The universal form factor $\tau_{1/2}^{ren}(y)$.
The curves refer to choices of the threshold parameters:
$\nu_c=2.0$ GeV,  $\nu^\prime_c=2.5$ GeV (continuous line),
$\nu_c=2.5$ GeV,  $\nu^\prime_c=3.0$ GeV (dashed line),
$\nu_c=3.0$ GeV,  $\nu^\prime_c=3.5$ GeV (dotted  line).
\label{tau12}}
\end{figure}

Considering again the three point correlator Eq.~(\ref{threep}) and using
$$J_1^{v^\prime}=J_s^{v^\prime}={\bar q} h^c_{v^\prime},\quad
J_W={\tilde A}_\mu={\bar h}^c_{v^\prime} \gamma_\mu \gamma_5 h^b_v,\quad 
{\rm and}\quad J_2^v=J_5^v={\bar q}i \gamma_5 h^b_v,$$ 
it  is possible to derive the sum rule for the form
factor $\tau_{1/2}^{ren}$. 
Also in this case the $\alpha_s$ correction in the perturbative term is
sizeable, but it turns out to be partially compensated by the analogous
correction in the leptonic constants $F$, $F^+$. Notice that this is a   
remarkable result, not expected 
{\em a priori}, since the normalization of the
form
factor, for example at zero recoil, is not fixed by symmetry arguments.
The perturbative corrections, however, do not equally affect the form
factor
for all values of the variable $y$, but they are sensibly $y$ dependent,
with the effect of increasing the slope of $\tau_{1/2}$ with respect to
the
case where they are omitted.
The result is depicted in Fig.~\ref{tau12},\cite{noitau} where the curves
refer to various choices for the continuum
thresholds. The region limited by the curves essentially determines the
theoretical accuracy of the calculation.
Considering the $y$ dependence, the limited range of values
allowed by the mass difference between $D$ and $D_0$ permits the expansion
near $y=1$,
\be
\tau_{1/2}^{ren}(y)=\tau_{1/2}(1)
\Big(1-\rho^2_{1/2} (y-1)+c_{1/2} (y-1)^2\Big) \;\;\;.
\ee
\noindent A three-parameter fit to Fig.~\ref{tau12} gives
\be   
\tau_{1/2}(1)=0.35\pm0.08 \;\;,
\hskip 1cm \rho^2_{1/2}=2.5\pm 1.0 \;\;, \hskip 1cm c_{1/2}=3\pm 3 \;\;.
\ee
An immediate application of this result concerns the prediction of the
semileptonic $B$ decay rates to $D_0$ and $D_1^\prime$.
Using
$V_{cb}=3.9 \times 10^{-2}$ and
$\tau(B)= 1.56 \times 10^{-12}$ sec, one obtains
\be
{\cal B}(B\to  D_0 \ell \bar \nu)=(5 \pm 3) \times 10^{-4},
\hspace{1cm}
{\cal B}(B\to D^\prime_1 \ell \bar \nu)= (7 \pm 5) \times 10^{-4} \;\;\;.
\ee
This means that only a very small fraction of the semileptonic $B \to X_c$
decays  is represented by transitions into the
$s_\ell^P={1\over 2}^+$ charmed doublet. 

\begin{figure}[htb]
\hskip 0.5in
\psfig{figure=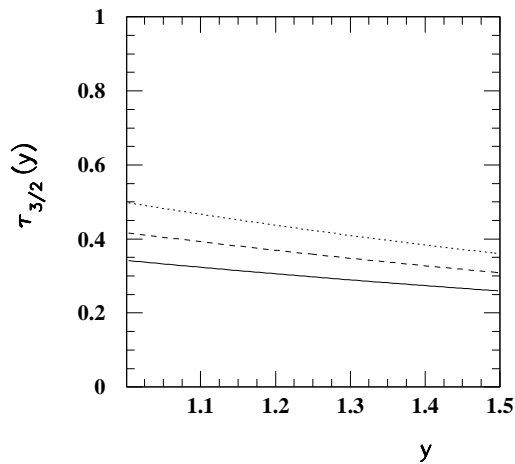,width=8cm}
\vspace{-1.cm}
\caption{The universal form factor $\tau_{3/2}(y)$ The three curves
refer to different choices of the threshold parameters in the
perturbative contribution to the sum rule. \label{tau32}}
\end{figure}

The $B$ meson transitions to final  charmed states belonging to the
$s_\ell=3/2$ doublet are described by the universal
form factor $\tau_{3/2}(y)$. This is defined by the following matrix
elements in the heavy quark limit:
\bea
&&\langle D_1(v^\prime,\epsilon)|{\bar h}^c_{v^\prime}\gamma_\mu
(1-\gamma_5) h^b_v|B(v)\rangle  = {\sqrt {m_B m_{D_1}}}   
 \hskip 4pt \tau_{3/2}(v\cdot v^\prime)  \nonumber  \\
&&\times \Big\{ [{ (1 - (v \cdot v^\prime)^2 ) \over \sqrt 2} \;\;
\epsilon^*_\mu  
+ { (\epsilon^* \cdot v) \over \sqrt 2} [- 3 \hskip 4pt v_\mu +
 (v \cdot v^\prime -2) \hskip 4pt v^\prime_\mu ]  \nonumber \\
&&+
i  {(v \cdot v\prime - 1) \over 2 \sqrt 2} \epsilon_{\mu\nu\alpha\beta}
\epsilon^{*\nu} (v+v^\prime)^\alpha \cdot 
  (v-v^\prime)^\beta] \Big\}  \; ,  \label{tau32a}
\eea
\bea
&&\langle D_2(v^\prime,\epsilon)|{\bar h}^c_{v^\prime}\gamma_\mu
(1-\gamma_5) h^b_v|B(v)\rangle  = {\sqrt
{m_B m_{D_2}}}
 \hskip 3pt \tau_{3/2}(v\cdot v^\prime)  \nonumber
\\
&&\times
[ \hskip 3pt i \hskip 4pt {{\sqrt 3} \over 2}
\epsilon_{\mu\alpha\beta\gamma} 
\epsilon^{*\alpha\nu} v_\nu (v+v^\prime)^\beta (v-v^\prime)^\gamma
\nonumber \\
&&- {\sqrt 3} (v \cdot v^\prime +1) \epsilon^*_{\mu\alpha} v^\alpha
+{\sqrt 3} \epsilon^*_{\alpha\beta} v^\alpha v^\beta
v^\prime_\mu ]. \label{tau32b} 
\eea
\noindent The QCD sum rule analysis for this function is available in
the leading order in $\alpha_s$.\cite{lorotau} The procedure closely follows
the one outlined above, with  the interpolating currents 
$$J_v^{\mu \nu}={1 \over \sqrt{6}}{\bar q}
(\stackrel{\leftrightarrow}{\partial^\mu}
\gamma^\nu
+\stackrel{\leftrightarrow} {\partial^\nu} \gamma^\mu -{1 \over 2} g^{\mu
\nu}\stackrel{\leftrightarrow}{\partial^\rho} \gamma_\rho)h_v$$ chosen 
to interpolate
the $D_2$ meson with $J^P=2^+$, and 
$${\tilde J}_v^\mu={\bar q} \gamma_5
\stackrel{\leftrightarrow}{\partial^\mu} h_v$$
interpolating the $D_1$ meson with $J^P=1^+$,
\bea
\langle 0|J_v^{\mu \nu}| D_2(p,\epsilon)\rangle  &=& \epsilon_{\mu \nu} F^+_{3/2}
\sqrt{m_c} \; , 
\nonumber \\[0.2cm]
\langle 0|{\tilde J}_v^\mu | D_1(p, \epsilon)\rangle  &=& \epsilon_\mu F^+_{3/2}
\sqrt{m_c} \;\;. \label{const32}
\eea
\noindent  The mass parameter analogous to those defined in
Eqs.~(\ref{hadmass}),~(\ref{lbarpiu}) is given by
\be
M_{D_2,D_1}=m_c+ {\bar \Lambda}_{3/2} \;\;. \label{lbar32}
\ee
\noindent The sum rule analysis for these quantities
gives$\,$\cite{lorotau}
\be
{\bar \Lambda}_{3/2}=1.05 \pm 0.10 \; {\rm GeV}\,, \qquad 
F^+_{3/2}=0.43 \pm 0.06 \; {\rm GeV}^{5/2} \;.
\ee
As for the form factor $\tau_{3/2}(y)$, 
 the result of the sum rule is shown in
Fig.~\ref{tau32}; it corresponds to 
\be
\tau_{3/2}(y)=\tau_{3/2}(1)[ 1-\rho^2_{3/2}(y-1)] \label{tau32fit} 
\ee
\noindent with $\tau_{3/2}(1)\simeq 0.4$ and $\rho^2_{3/2}\simeq 0.52$.
These values lead to  the  predictions
\be
{\cal B}(B\to  D_1 \ell \bar \nu)\simeq 3.2 \times 10^{-3}\,,
\hspace{0.7cm}
{\cal B}(B\to D_2 \ell \bar \nu)\simeq 4.8 \times 10^{-3} \;\;\;.
\ee

Other determinations of the $\tau$ functions have appeared in the
literature, employing various versions of the constituent quark
model.\cite{tauqm} The results range in quite a large interval,
$\tau_{1/2}(1)=0.06-0.40$, $\rho_{1/2}^2=0.7-1.0$ and
$\tau_{3/2}(1)=0.31-0.66$, $\rho^2_{3/2}=1.4-2.8$. These results
critically depend on the features of the models employed.    

Finally, let us mention that  the universal form factors describing
the $B$ meson transitions to the higher doublets with 
$s_\ell^P={3 \over 2}^-$ and $s_\ell^P={5 \over 2}^-$ have 
also been computed
by QCD sum rules.\cite{noi52} Such doublets comprise the states with
$J^P_{s_\ell}=(1^-,2^-)_{3/2}$ and 
$J^P_{s_\ell}=(2^-,3^-)_{3/2}$, respectively.
In this case, the application of the method to high-spin states shows
several difficulties, mainly due to the identification of the range of the
parameters needed in the sum rule analyses, because of the particular
features of the considered states and of their interpolating
currents. Some information could be obtained from other theoretical
approaches, namely constituent quark models predicting the heavy meson
spectrum. The final result, although affected by a sizeable theoretical
uncertainty, nevertheless is useful for assessing the role of high-spin
meson doublets in constituting a part of the charm inclusive semileptonic
$B$ decay width. Denoting the members of the considered doublets by
$(D^*_1,D^*_2)$, $(D^{* \prime}_2, D^*_3)$  corresponding 
to $s_\ell^P={3/2}^-$ and $s_\ell^P={5/2}^-$, respectively,
one finds that~\cite{noi52}
\be
{\cal B}(B \to D_2^{* \prime} \ell \nu_\ell) \simeq 
{\cal B}(B \to D^*_3\ell \nu_\ell) \simeq 1 \times 10^{-5} \;\;.
\ee
\noindent On the other hand, the $B$ decays to the $s_\ell^P={3/2}^-$
doublet turn out to be negligible at the leading order in the
${1 / m_Q}$ expansion, due to the small value of the
corresponding universal form factor.

At  the end of our discussion about the universal form factors describing
$B$ decays in the $m_Q \to \infty$ limit, it should be stressed that an
important issue concerns the role of ${1 / m_Q}$
corrections, which introduce subleading form factors. The analysis of such
corrections has been carried out for a number of relevant cases, and we
refer the reader to the relevant literature.\cite{neubpr,daimq,stewart}

\section{Inclusive Heavy Hadron Decays}
We have outlined above how the use of heavy quark symmetries allows a
simplified description of exclusive heavy hadron decays, both introducing
universal form factors and  allowing to classify the symmetry-breaking
contributions using an expansion in the inverse heavy quark
mass.\cite{ural} 
The presence of a large mass 
also turns out to be very useful when dealing
with  inclusive decays, i.e. 
the transitions of a given particle into all possible final states with
assigned quantum numbers.
Such decays
represent theoretically ``clean" processes, since it is expected that in
the sum over many hadrons the bound state effects, which are the main
source of uncertainty in the calculation of exclusive transitions, should
be eliminated by the averaging procedure. This expectation is based on the 
quark-hadron duality.\cite{misha} Besides, the possibility of computing
inclusive decay rates of heavy hadrons by means of an expansion in the
inverse powers of the heavy quark mass provides in this case a
reliable QCD-based systematic approach.

Let us
briefly summarize the main aspects of such  QCD analysis of the inclusive
decay widths of the heavy hadrons.
The starting point is the transition operator 
$\hat T(Q \to X_f \to Q)$,\cite{old}
\begin{equation}
\hat T=i \int d^4 x \; T[{\cal L}_W(x){\cal L}_W^\dagger(0)] 
\label{t} 
\end{equation}
\noindent 
describing an amplitude with the heavy quark $Q$
having the  same momentum in the initial and final state.
${\cal L}_W$ is the effective weak Lagrangian governing the decay
$Q \to X_f$.
The inclusive width of the hadron $H_Q$ can be obtained by 
averaging $\hat T$ over $H_Q$ and taking the imaginary part of the forward 
matrix element,
\begin{equation}
\Gamma(H_Q \to X_f)={ 2 \; Im~\langle H_Q|\hat T|H_Q\rangle  \over 2\; M_{H_Q}} \hskip 3 pt.
\label{width} 
\end{equation}
\noindent 
The reason behind calculating the right-hand side
 of Eq.~(\ref{width})
is to set up an operator product expansion  
for the transition operator $\hat T$ in terms of local operators 
${\cal O}_i$,
\begin{equation}
\hat T= \sum_i C_i {\cal O}_i \;\;\; \label{ope}
\end{equation}
with ${\cal O}_i$ ordered according to their dimension, and the 
coefficients $C_i$ containing appropriate inverse powers of the heavy quark 
mass $m_Q$.
The lowest dimension operator appearing in Eq.~(\ref{ope}) is ${\cal
O}_3=\bar Q Q$.
The next gauge and Lorentz invariant operator is the $D=5$ chromomagnetic 
operator
${\cal O}_G= {\bar Q} {g \over 2}  \sigma_{\mu \nu} G^{\mu \nu} Q$.

The matrix element of $\bar Q Q$ over $H_Q$ can be obtained using the heavy 
quark equation of motion, expanded in the heavy quark mass,
\begin{equation}
\bar Q Q = \bar Q \gamma^0 Q + { {\cal O}_G \over 2 m_Q^2} -
{ {\cal O}_\pi \over 2 m_Q^2} + \, O(m_Q^{-3}) \,,
\end{equation}
where ${\cal O}_\pi$ is the kinetic energy operator 
${\cal O}_\pi={\bar Q} (i {\vec D})^2 Q$.
On the other hand,  the $H_Q$ matrix element  of $\bar Q \gamma^0 Q$ 
is unity (modulo the covariant normalization of the states).

The number of independent operators appearing in Eq.~(\ref{ope}) increases
if the
 $1/m_Q^3$ term is considered. Such operators are of the four-quark type
\begin{equation}
{\cal O}_6^q = {\bar Q} \Gamma q \; {\bar q } \Gamma Q \label{4q}
\end{equation}
where $\Gamma$ is an appropriate combination of the Dirac and color matrices.

In this way, a complete classification of  various contributions
to the inclusive decay rates can be obtained for  different hadrons $H_Q$.
In the expression for the inclusive width $\Gamma(H_Q \to X_f)$
\begin{equation}
\Gamma(H_Q \to X_f)=\Gamma_0^f \; \Big[ A_0^f 
+{A_2^f \over m_Q^2} +{A_3^f \over m_Q^3} +
...\Big] \label{expan}
\end{equation}
the $A_i^f$ factors which (together with $\Gamma_0^f$) depend on the 
final state $X_f$,  
include perturbative short-distance coefficients and  
nonperturbative hadronic matrix elements incorporating the
long range dynamics.
The partonic prediction for the width in Eq.~(\ref{expan})
corresponds to the leading term 
$\Gamma^{part}(H_Q \to X_f)=\Gamma_0^f A_0^f$, with $A_0^f=1+ c^f {\alpha_s 
/ \pi}+ O(\alpha_s^2)$ and $\Gamma_0^f \propto m_Q^5$; 
differences among the widths of the  
hadrons $H_Q$  emerge at the next to leading order in $1/m_Q$ and are related 
to  different values of the matrix elements of the operators ${\cal O}_i$
of dimension larger than three.

It is important to notice the absence of 
the first order term $m_Q^{-1}$ in Eq.~(\ref{expan}), 
a result obtained by
Chay, Georgi and Grinstein,\cite{chay} and Bigi, Uraltsev and
Vainshtein.\cite{buv} This depends on the adopted choice of the definition
of the heavy quark mass, as explained in Sec.~\ref{hqt}.

The occurrence of operators of the type (\ref{4q}) 
is an appealing feature of the expansion in
Eq.~(\ref{ope}), as far as the determination of the inclusive widths
is concerned. As a matter of 
fact, contrary to the $D=5$ operators ${\cal O}_G$ and ${\cal O}_\pi$ 
which are spectator 
blind, the $D=6$ operators give different contributions when averaged over
hadrons belonging to the same $SU(3)$ light flavor multiplet, and therefore
they are responsible of  different lifetimes  
of, e.g., $B^-$ and $B_s$, $\Lambda_b$ and $\Xi_b$.
The spectator flavor dependence is related to the mechanisms of weak 
scattering and Pauli interference,\cite{old} both suppressed by the
factor 
$m_Q^{-3}$ with respect to the parton decay rate.

In the following subsection, we shall see how the $1 /
m_Q$ expansion
provides a systematic framework to discuss one of the open questions in
the $b$ phenomenology, i.e. the problem of the ratios of beauty
hadron lifetimes.

\subsection{The Problem of Beauty Hadron Lifetime Ratios}
An interesting problem of  present-day heavy quark physics is
represented by
the measured difference between the
$\Lambda_b$ baryon and $B_d$ meson lifetimes,
$\tau(\Lambda_b^0)= 1.208 \pm 0.051 \; ps$ and
$\tau(\bar {B^0})= 1.548 \pm 0.021 \; ps$.\cite{lephfs}
As a matter of fact, the deviation from
unity, at the level of $20 \%$, of the ratio
$\tau(\Lambda_b)/\tau(B_d)$, 
$\tau(\Lambda_b^0)/\tau(\bar {B^0}) = 0.780 \pm 0.037$,\footnote{These
data were reported by the Particle Data Group on the basis of the
analyses reported by the LEP $B$ Lifetime Group.\cite{lephfs}} is in
contradiction with the naive
expectation that, at the scale of the $b$ quark mass, the spectator model
should describe rather accurately the decays of the hadrons containing one
heavy quark. 
The ratio $\tau(\Lambda_b)/\tau(B_d)$ can be computed in QCD using the
previously described approach. We have already noticed that the first term
of the expansion in Eq.~(\ref{expan}) reproduces the parton model result,
and
therefore it contributes universally to the inclusive decay width of all
hadrons containing the same heavy quark. Besides, we also pointed out that
the flavor of the spectator quark is first felt at ${\cal O}(m_Q^{-3})$. 

As for the differences in the lifetime of mesons and baryons, they could
already arise at the level $m_Q^{-2}$, both due  to the chromomagnetic
contribution and to the kinetic energy term in Eq.~(\ref{width}). 
In particular, the kinetic energy term is responsible for the 
difference for systems where the chromomagnetic contribution 
vanishes, namely in the case of $\Lambda_b$ and $\Xi_b$  
(the light degrees of freedom in $S$ wave). 
However, the results of a
calculation of $\mu^2_\pi$ for mesons$\;$\cite{braun}
and baryons$\;$\cite{lorolam} support the conjecture$\;$\cite{wise}
that the kinetic energy operator has practically 
the same matrix element when computed on such hadronic systems. 
The approximate equality of the kinetic energy operator on $B_d$ 
and $\Lambda_b$ can also be inferred by  considering that,
to the leading order in $1/m_Q$, 
$\mu^2_\pi(\Lambda_b)$ can be related~\cite{bigi2} to  $\mu^2_\pi(B_d)$ and
to the heavy quark masses by an expression which assumes the 
charm mass $m_c$ heavy enough for a meaningful
expansion in $1/m_c$. Namely, 
\bea
\mu^2_\pi(\Lambda_b)- \mu^2_\pi(B_d)&\simeq &
\frac{m_b m_c}{2 (m_b-m_c)}  \label{diff} \\
&\times& \left[\left(M_B+3 M_{B^*}-4 M_{\Lambda_b}\right)-
\left(M_D+3 M_{D^*}-4 M_{\Lambda_c}\right)\right].\nonumber
\eea
Using present data and the CDF measurement
$M_{\Lambda_b} = 5623 \pm 5 \pm 4$ MeV~\cite{kroll}
Eq.~(\ref{diff}) gives 
$\mu^2_\pi(\Lambda_b)- \mu^2_\pi(B_d) \simeq
0.002\pm 0.024~{\rm GeV}^2$, where the error mainly comes from the error in 
$M_{\Lambda_b}$.
The QCD sum rule outcome for 
$\mu^2_\pi(H_Q)$ is
$\mu^2_\pi(B_d) \simeq \mu^2_\pi(\Lambda_b) \simeq 0.6 \; {\rm GeV}^2$,
with an estimated uncertainty of about $30 \%$.\cite{braun,lorolam}
This result implies that the 
differences between the meson and baryon lifetimes should  
occur at the $m_Q^{-3}$ level, thus involving the four-quark operators
in Eq.~(\ref{4q}). They can be classified as follows:\cite{sachrajda}
\begin{eqnarray}
O^q_{V-A}&=&{\bar Q}_L \gamma_\mu q_L \; {\bar q}_L \gamma_\mu Q_L  \; , \nonumber \\
O^q_{S-P}&=&{\bar Q}_R q_L \; {\bar q}_L  Q_R  \; , \nonumber \\
T^q_{V-A}&=&{\bar Q}_L \gamma_\mu {\lambda^a \over 2} q_L \;
{\bar q}_L \gamma_\mu {\lambda^a \over 2} Q_L  \; , \nonumber \\
T^q_{S-P}&=&{\bar Q}_R {\lambda^a \over 2} q_L \; {\bar q}_L  
{\lambda^a \over 2} Q_R  \; , \label{4q1}
\end{eqnarray}
\noindent with $\displaystyle{q_{R,L}= {1 \pm \gamma_5 \over 2} q}$ and
$\lambda_a$ the Gell-Mann  matrices.

For mesons, the vacuum saturation approximation can be used 
to compute the matrix elements of the operators in Eq.~(\ref{4q1}),
\begin{eqnarray}
\langle B_q | O^q_{V-A} | B_q\rangle _{VSA} &=& \Big({m_b + m_q \over M_{B_q}} \Big)^2 
\langle B_q | O^q_{S-P} | B_q\rangle _{VSA} = {f_{B_q}^2 M_{B_q}^2 \over 4}  \; , 
\nonumber\\
\langle B_q | T^q_{V-A} | B_q\rangle _{VSA} &=& 
\langle B_q | T^q_{S-P} | B_q\rangle _{VSA} = 0 \;\;\; .\label{vsa}
\end{eqnarray}
\noindent Therefore, the matrix elements are expressed
in terms of quantities such as $f_B$ and the quark masses, and the 
resulting numerical 
values can be used in the calculation of the lifetimes, with the only caveat
concerning the accuracy
of the factorization approximation.\cite{sachrajda} 

The vacuum saturation approach cannot be employed for baryons;
in this case a direct calculation of the matrix elements is required,
for example using constituent quark models. 

A simplification can be obtained$\;$\cite{sachrajda} 
for $\Lambda_b$ using color and Fierz identities and 
introducing the operators
\begin{equation}
\tilde{\cal O}^q_{V-A} = {\bar Q}^i_L \gamma_\mu Q^i_L \; 
{\bar q}^j_L \gamma^\mu q^j_L \label{otilde}
\end{equation}
and
\begin{equation}
\tilde{\cal O}^q_{S-P} = {\bar Q}^i_L  q^j_R \; 
{\bar q}^j_L Q^i_R 
\end{equation}
($i$ and $j$ are color indices). As a matter of fact, 
the $\Lambda_b$ matrix elements of the  operators in Eq.~(\ref{4q1}) 
can be expressed in terms of 
$\langle \Lambda_b | \tilde{\cal O}^q_{V-A}|\Lambda_b\rangle $ and
$\langle \Lambda_b | {O}^q_{V-A}|\Lambda_b\rangle $, modulo $1/m_Q$ corrections 
contributing to subleading terms in the expression for the inclusive widths.

The matrix element of 
$\tilde {\cal O}^q_{V-A}$ and ${\cal O}^q_{V-A}$ can be parametrized as
\begin{equation}
\langle \tilde {\cal O}^q_{V-A}\rangle_{\Lambda_b} =
{\langle \Lambda_b | \tilde {\cal O}^q_{V-A} |\Lambda_b\rangle  \over 2 M_{\Lambda_b} } =
{f_B^2 M_B \over 48} r  \label{par}
\end{equation}
and 
\begin{equation}
{\langle \Lambda_b | {O}^q_{V-A} |\Lambda_b\rangle   } = - {\tilde B} \;\;
{\langle \Lambda_b | \tilde {\cal O}^q_{V-A} |\Lambda_b\rangle   } \label{btilde}
\end{equation}
with $\tilde B=1$  in the valence quark approximation.

For $f_B=200$ MeV and $r=1$, Eq.~(\ref{par}) corresponds to the value: 
$\langle \tilde {\cal O}^q_{V-A}\rangle_{\Lambda_b} = 4.4 \times 10^{-3} \;
{\rm GeV}^3$.
The $\Lambda_c$ matrix element of
$\tilde {\cal O}^q_{V-A}$ has been computed using a bag model and a
nonrelativistic quark model;\cite{bilic} the results
$\langle \tilde {\cal O}^q_{V-A}\rangle_{\Lambda_c} \simeq 0.75 \times 10^{-3}
\; {\rm GeV}^3$ and
$\langle \tilde {\cal O}^q_{V-A}\rangle_{\Lambda_c} \simeq 2.5 \times 10^{-3} 
\; {\rm GeV}^3$, correspond to
$r \simeq 0.2$ and $r \simeq 0.6$, respectively.
A different model$\;$\cite{der} is used in  another analysis.\cite{rev}

Larger values of the matrix elements
 have been advocated by Rosner$\;$\cite{rosner}
using the values of the mass splitting $\Sigma_b^* - \Sigma_b$ and
$\Sigma_c^* - \Sigma_c$, and assuming that the $\Lambda_b$ and $\Sigma_b$ 
wave functions are similar, 
$r \simeq 0.9 \pm 0.1$, taking 
$M^2_{\Sigma_b^*} - M^2_{\Sigma_b}=M^2_{\Sigma_c^*} - M^2_{\Sigma_c}$, 
or $r \simeq 1.8 \pm 0.5$ 
using the DELPHI measurement.\cite{delphi}

Information from constituent quark models have been supplemented 
by estimates based on field theoretical approaches, such as 
QCD sum rules and lattice. As a matter of fact, a large value of $r$,
namely $r\simeq 4\;-\;5$, 
would explain the difference between $\tau(\Lambda_b)$ 
and $\tau(B_d)$.\cite{sachrajda}
The application of the QCD sum rule method to the calculation 
of the matrix element of an operator of high dimension
presents a number of disadvantages; 
nevertheless, interesting and quite reliable information can be
obtained.  The result$\;$\cite{noilam} is
$\langle \Lambda_b | \tilde {\cal O}^q_{V-A} |\Lambda_b\rangle \simeq
(0.4-1.20) \times 10^{-3}\; {\rm GeV}^3$, corresponding to the
parameter $r$ in the range  
$r \simeq 0.1-0.3$.\footnote{Another QCD sum rule 
analysis~\cite{cinesi} finds a
larger value for the parameter $r$. However, this result strongly depends
on the  assumption of a huge deviation from the vacuum saturation
approximation for the four-quark condensate.}
 The conclusion of such
analysis is that the inclusion of $1/m_Q^3$ terms in the expression of the
inclusive widths does not solve the puzzle represented by the
difference between
$\tau(\Lambda_b)$ and $\tau(B_d)$. As a matter of fact,
using the formulae   for the lifetime ratio in the
framework of the heavy quark expansion,\cite{sachrajda}
the QCD sum rule result  together with $\tilde B=1$ gives
\be
\tau(\Lambda_b)/\tau(B_d) \ge 0.94 \; \; \;.
\ee  
 \noindent As for lattice QCD, a  calculation
with static $b$ quark
finds higher values for $r$, $r \simeq 1.2 \pm 0.2$, which nevertheless is
not enough to explain the observed discrepancy in the 
lifetime ratio.\cite{lattice4q} 

In conclusion, the expansion in the inverse  heavy quark mass provides a
systematic framework to study inclusive heavy hadron decays. In
particular it allows to compute the corrections to the partonic picture,
which identifies with the first term of such 
an expansion, in terms of the
matrix elements of suitable operators, weighted by increasing powers of
$m_Q^{-1}$. However, the application of such a formalism to the
computation of the lifetime ratio ${\tau(\Lambda_b) /
\tau(B_d)}$ does not explain yet the observed ratio, leaving this issue as
one of the exciting open problems in the phenomenology of beauty
hadrons.

\section{Conclusions and Perspectives}
Weak decays of heavy quarks are especially appealing, mainly because the
presence of a large scale, i.e. $m_Q$, allows  to formulate model
independent relations among nonperturbative quantities which are
difficult to access in full QCD. We have shown what we consider among 
the most
relevant phenomenological examples: the use of heavy quark
symmetry in parametrizing weak matrix elements between heavy hadrons in
terms of universal form factors, and the application of 
${1 / m_Q}$ expansion to compute reliably inclusive
decay rates of heavy
hadrons. In both cases, the most straightforward application is the
extraction of the CKM matrix elements $V_{cb}$ and $V_{ub}$. The former
has been  determined with increasing accuracy, the latter, being
intrinsically smaller, is still a subject of ongoing analyses.

Many interesting topics have not been touched upon. The list includes rare
$B$ decays, among which those induced by flavor changing neutral current
represent a fertile ground for exploring physics beyond the Standard
Model; these are commonly viewed as those processes which can tell us
something about new physics before the LHC era. A remarkable example are
radiative decays induced by the $b \to s$ transition, measured both in the
inclusive and exclusive $B \to K^* \gamma$
modes,\cite{pdg,babarrad}
which  put significant constraint on the space of parameters of
various new physics scenarios. 

Another untouched sector concerns nonleptonic decays, where the impact of
nonperturbative QCD effects is the highest. The effective Hamiltonian we have
given in Sec.~\ref{hamilt} is the starting point for the
theoretical treatment of these processes. It allows us to sort out the 
nonperturbative quantities, the hadronic matrix elements of the operators in
the OPE, from the perturbative ones, i.e. the Wilson coefficients. Having
identified them, the problem of giving a reliable estimate of such
quantities is still unsolved. Usually, either models or simplified
approaches have been employed, among which a widely used one
is factorization, either in its simpler formulation, {\it naive}
factorization,\cite{bsw} or in its most refined versions.\cite{stech}
This is quite a general problem for all nonleptonic decays, 
which is not specific of heavy hadrons. Nevertheless, the
situation seems once more easier in this case, since it turns out that
factorization works correctly in the heavy quark limit, at least for those
decays usually classified as {\it Class I}.\cite{buchalla} This is
probably (and hopefully) a field in growth.

Nonleptonic decays play a central role in the analyses of CP violation
in $B$ decays, since the most promising channels to extract the angles
of the unitarity triangle are just nonleptonic processes. The issue of CP
violation is one of the most extensively reviewed, and we refer the reader
to the existing literature.\cite{babar}

Finally, I have not discussed the $B_s$ and $B_c$ decays, which will
play a role at the CERN Large Hadron Collider (LHC).\cite{report} 

It seems that in the very last years the ongoing experiments
in high-energy physics  are providing us
with many exciting new results, from the SuperKamiokande observation of
the atmospheric neutrino anomaly,\cite{sk} to
the new measurements~\cite{eps} of ${\rm Re}(\epsilon^\prime/\epsilon )$, to
the very recent direct observation of the $\nu_\tau$ in July 2000.\cite{donut} 

As for $B$ physics, the $B$-factories have
provided us with the very first results in summer 2000, thus opening the
season of intense investigations in this sector of the elementary
interactions. In a few years, the advent of LHC will tell us even more. 
The heavy hadron physics presents a number of exciting
perspectives as well as intriguing problems still to be solved, such as
the one of the beauty hadron lifetime ratios. 
Moreover, the precise measurements of CP violating processes and of the
phases of the CKM matrix elements will provide us with many answers and,
at the same time, will open new problems.

In 1990, J.D. Bjorken~\cite{bj90} pointed out that there was a great
potential in the
discovery of new symmetries in the heavy quark sector. I hope to have
shown that he was right.

\section*{Acknowledgments}
I warmly thank M. Shifman for inviting me to join his effort to write this
{\em Handbook of QCD}. 
I also would like to thank all those with whom I had the pleasure to work,
who all had an important impact on my professional growth: P. Colangelo,
R. Gatto, G. Nardulli, M. Neubert,  N. Paver and M.R. Pennington. Many
useful discussions
with G. Buchalla and A. Khodjamirian are acknowledged.

I am grateful for support from the EU-TMR Programme, Contract
No. CT98-0169, EuroDA$\Phi$NE.


\section*{References}
\begin{thebibliography}{99}

\bibitem{sak}A. Sakharov, {\em JETP Lett.} {\bf 5}, 24 (1967).

\bibitem{isw}N. Isgur and M.B. Wise, \Journal{\PLB}{232}{113}{1989}; 
\Journal{\PLB}{237}{527}{1990}. 

\bibitem{eichten}E. Eichten and B. Hill,
\Journal{\PLB}{234}{511}{1990}.

\bibitem{georgi}H. Georgi, \Journal{\PLB}{240}{447}{1990}.

\bibitem{shuryak}E. Shuryak, \Journal{\NPB}{198}{83}{1982}; \\
S. Nussinov and W. Wetzel, \Journal{\PRD}{36}{130}{1987}.

\bibitem{hqet_1}M. Voloshin and M. Shifman,
{\em Sov. J. Nucl. Phys. }{\bf 47}, 511 (1988).

\bibitem{uraltsev}N. Uraltsev, this Volume.

\bibitem{shifman} M.A. Shifman, A.I. Vainshtein and V.I. Zakharov,
\Journal{\NPB}{147}{385}{1979}.\\
 For a review on the QCD sum rule
method see the reprint volume {\it Vacuum Structure and QCD Sum Rules},
ed. M.A. Shifman (North-Holland, Amsterdam, 1992).

\bibitem{col}P. Colangelo and A. Khodjamirian, this Volume.

\bibitem{wilson}K.G. Wilson, \Journal{\PRD}{179}{1499}{1969}.

\bibitem{buras}For a review see: 
G. Buchalla, A.J. Buras and M.E. Lautenbacher, {\em Rev. Mod. Phys.}
{\bf 68}, 1125 (1996). 

\bibitem{neubpr}A  comprehensive review of the method can be found in
M. Neubert, {\em Phys. Rept.} {\bf 245}, 259 (1994). 


\bibitem{mu2pi}
V. Chernyak, \Journal{\PLB}{387}{173}{1996}; \\
M. Gremm, A. Kapustin, Z. Ligeti and M.B. Wise,
\Journal{\PRL}{77}{20}{1996}; \\
P. Ball and V. Braun, \Journal{\PRD}{49}{2472}{1994};\\
E. Bagan, P. Ball, V. Braun and P. Gosdzinsky,
\Journal{\PLB}{342}{362}{1995};\\
M. Neubert, \Journal{\PLB}{389}{727}{1996}; \\
F. De Fazio, {\em Mod. Phys. Lett.} {\bf A 11}, 2693 (1996); \\
D.S. Hwang, C.S. Kim and
W. Namgung, \Journal{\PLB}{406}{117}{1997};\\
 V. Gimenez, G. Martinelli and
C.T. Sachrajda, \Journal{\NPB}{486}{227}{1997}.

\bibitem{grin}
H. Georgi, B. Grinstein and M.B. Wise, \Journal{\PLB}{252}{456}{1990}.

\bibitem{renormalon}
I.I. Bigi and N.G. Uraltsev, \Journal{\PLB}{321}{412}{1994};\\
I.I. Bigi, M.A. Shifman, N.G. Uraltsev and A. I. Vainshtein,
\Journal{\PRD}{50}{2234}{1994}; \\
M. Beneke and V. Braun, \Journal{\NPB}{426}{301}{1994}.

\bibitem{georgirev}A nice derivation of the properties of the Isgur-Wise
function can be found in H. Georgi, Boulder TASI {\bf 91}, 0589.


 \bibitem{pdg} D.E. Groom {\em et al.}, Particle Data Group, {\em
Eur. Phys. J.} {\bf C 15}, 1 (2000).

\bibitem{anderson}
S. Anderson {\it et al.}, CLEO Collab., {\em Nucl. Phys.} 
{\bf A 663}, 647 (2000).


\bibitem{falk92}A.F. Falk and M. Luke, \Journal{\PLB}{292}{119}{1992}.

\bibitem{colangelo95} P. Colangelo {\it et al.}, \Journal{\PRD}{52}{6422}
{1995};\\
 P. Colangelo and F. De Fazio, {\em Eur. Phys. J.} {\bf C 4}, 503 (1998).

\bibitem{kilian92}
U. Kilian, J.G. K\"orner and D. Pirjol,\Journal{\PLB}{288}{360}{1992}.

\bibitem{lep}P. Abreu {\it et al.}, DELPHI Collab.,
\Journal{\PLB}{345}{598}{1995};\\
R. Akers {\it et al.}, OPAL Collab., \Journal{\ZPC}{66}{19}{1995};\\
D. Buskulic {\em et al.}, ALEPH Collab., \Journal{\ZPC}{69}{393}{1996}.

\bibitem{ciulli}
V. Ciulli, hep-ex/9911044.

\bibitem{casal}R. Casalbuoni {\it et al.}, {\em Phys. Rep.} {\bf 281}, 145
(1997).

\bibitem{falk0}A.F. Falk, \Journal{\PLB}{305}{268}{1993}.

\bibitem{ademollo}
M. Ademollo and R. Gatto, \Journal{\PRL}{13}{264}{1964}.

\bibitem{roos}
H. Leutwyler and M. Roos, \Journal{\ZPC}{25}{91}{1984}.

\bibitem{bmunugamma}
D. Atwood, G. Eilam and A. Soni {\em Mod. Phys. Lett.}
{\bf A 11}, 1061 (1996);\\
G. Burdman, T. Goldman and D. Wyler, \Journal{\PRD}{51}{111}{1995};\\ 
P. Colangelo, F. De Fazio and G. Nardulli, 
\Journal{\PLB}{372}{331}{1996}; 
\Journal{\PLB}{386}{328}{1996}.

\bibitem{politzer}
H.D. Politzer and M.B. Wise,
\Journal{\PLB}{206}{681}{1988};\\
 \Journal{\PLB}{208}{504}{1988}.

\bibitem{neub1076}
M. Neubert, \Journal{\PRD}{46}{1076}{1992}.

\bibitem{ji}
X. Ji and M.J. Musolf, \Journal{\PLB}{ 257}{409}{1991}.

\bibitem{falk90}
A.F. Falk, H. Georgi, B. Grinstein and M.B. Wise,
\Journal{\NPB}{343}{1}{1990}.

\bibitem{neubert92}
M. Neubert, \Journal{\PRD}{45}{2451}{1992}.

\bibitem{kor} G.P. Korchemsky and A.V. Radyushkin,
\Journal{\NPB}{283}{342}{1987};\\
G.P. Korchemsky, {\em Mod. Phys. Lett.} {\bf A 4}, 1257 (1989).


\bibitem{neub92}
M. Neubert, \Journal{\PRD}{46}{2212}{1992}.

\bibitem{luke}
M.E. Luke, \Journal{\PLB}{252}{447}{1990}.

\bibitem{rieckert}
M. Neubert and V. Rieckert,  \Journal{\NPB}{382}{97}{1992}.

\bibitem{neub91}
M. Neubert, \Journal{\PLB}{264}{455}{1991}.

\bibitem{vcbwg}
The LEP $V_{cb}$ Working Group, http://lepvcb.web.cern.ch/LEPVCB.

\bibitem{rad}
A.V. Radyushkin, \Journal{\PLB}{271}{218}{1991}.

\bibitem{neub93}
M. Neubert, \Journal{\PRD}{47}{4063}{1993}.

\bibitem{lorotau}
P. Colangelo, G. Nardulli and N. Paver, \Journal{\PLB}{293}{207}{1992}.

\bibitem{noitau}
P. Colangelo, F. De Fazio and N. Paver, \Journal{\PRD}{58}{116005}{1998}.


\bibitem{dai}Y.H. Dai, C.S. Huang, M.Q. Huang and C. Liu,
\Journal{\PLB}{390}{350}{1997};\\
Y.H. Dai, C.S. Huang and   M.Q. Huang, \Journal{\PRD}{55}{5719}{1997}.

\bibitem{misha} The issue of quark-hadron duality is analyzed by
M.A. Shifman, this Volume.

\bibitem{matthias92}
M. Neubert, \Journal{\PRD}{46}{1076}{1992}

\bibitem{broad92}
D.J. Broadhurst and A.G. Grozin, \Journal{\PLB}{274}{421}{1992}.
   
\bibitem{bagan92}
E. Bagan, P. Ball, V.M. Braun and H. G. Dosch,
 \Journal{\PLB}{278}{457}{1992}.

\bibitem{kotikov}
A.V. Kotikov, \Journal{\PLB}{254}{158}{1991};
\Journal{\PLB}{259}{314}{1991}.

\bibitem{neubtasi}
M. Neubert, hep-ph/9404296, Lectures presented at TASI-93,
      Boulder, Colorado, 1993.


\bibitem{blok}
B. Blok and M. Shifman, \Journal{\PRD}{47}{2949}{1993}.

\bibitem{ball}
E. Bagan, P. Ball and P. Gosdzinsky, \Journal{\PLB}{301}{249}{1993}.

\bibitem{bj90}
J.D. Bjorken, Proceedings of the {\em 4th Rencontres de Physique
de la Vall\'ee d'Aoste}, La Thuile, Italy, 1990, ed. M. Greco,
(Editions
Fronti\`eres, Gif sur Yvette, 1990) pag. 583.

\bibitem{vol92}
M.B. Voloshin, \Journal{\PRD}{46}{3062}{1992}.

\bibitem{ural}
For a review see: I. Bigi, M. Shifman and N.G. Uraltsev,
{\em Ann. Rev. Nucl. Part. Sci.} {\bf 47}, 591  (1997).


\bibitem{isgur91}   
N. Isgur and M.B. Wise, \Journal{\PRD}{43}{819}{1991}.

\bibitem{poling}
R.A. Poling, Talk given at the {\em 19th International Symposium on Lepton
and Photon Interactions at High-Energies} (LP 99), Stanford, California,
9-14 Aug. 1999, hep-ex/0003025.

\bibitem{nonlepb}
P. Colangelo, F. De Fazio and G. Nardulli,
\Journal{\PLB}{303}{152}{1993};\\
G. Lopez Castro and J. H. Munoz, \Journal{\PRD}{55}{5581}{1997};\\
M. Neubert, \Journal{\PLB}{418}{173}{1998}.


\bibitem{pene98}
J. Charles, A. Le Yaouanc, L. Oliver, O. Pene and J.C. Raynal,
\Journal{\PLB}{425}{375}{1998};
{\it ibidem} {\bf 433}, 441 (1998) (E);\\
P. Colangelo, F. De Fazio, G. Nardulli and
N. Paver, \Journal{\PRD}{60}{033002}{1999}. 

\bibitem{dai1}Y.B. Dai, C.S. Huang, M.Q. Huang, H.Y. Jin and C. Liu, 
\Journal{\PRD}{58}{094032}{1998}; {\it ibidem}
{\bf D 59}, 059901 (1999) (E). 


\bibitem{tauqm}T.B. Suzuki, T. Ito, S. Sawada and M. Matsuda,
{\em Prog. Theor. Phys.} {\bf 91}, 757 (1994);\\
 A. Wambach, \Journal{\NPB}{434}{647}{1995};\\
 S. Veseli and
M.G. Olsson, \Journal{\PLB}{367}{302}{1996};
\Journal{\ZPC}{71}{287}{1996};\Journal{\PRD}{54}{886}{1996};\\
V. Morenas {\it et al.}, \Journal{\PRD}{56}{5668}{1997};\\
 A. Deandrea {\it et al.}, \Journal{\PRD}{58}{034004}{1998};\\
D. Ebert, R.N. Faustov and V.O. Galkin, \Journal{\PLB}{434}{365}{1998}.

\bibitem{noi52}
P. Colangelo, F. De Fazio and G. Nardulli, \Journal{\PLB}{478}{408}{2000}.

\bibitem{daimq}M. Huang, C. Li and Y.B. Dai,
\Journal{\PRD}{61}{054010}{2000}. 

\bibitem{stewart}A.K. Leibovich, Z. Ligeti, I.W. Stewart and M.B. Wise, 
\Journal{\PRL}{78}{3995}{1997};
\Journal{\PRD}{57}{308}{1998}.

\bibitem{old}
M.A. Shifman and M. Voloshin in the review by V. Khoze and M.A. Shifman,
{ \em Sov. Phys. Usp.} {\bf 26}, 387 (1983);\\
N. Bili\'c, B. Guberina and J. Trampetic,
\Journal{\NPB}{248}{261}{1984};\\
M. Voloshin and M.A. Shifman, {\em Sov. J. Nucl. Phys.} {\bf 41}, 120  
(1985);
{\em Sov. Phys. JETP} {\bf 64}, 698 (1986).

\bibitem{chay}
J. Chay, H. Georgi and B. Grinstein, \Journal{\PLB}{247}{399}{1990}.

\bibitem{buv}
I. Bigi, N. Uraltsev and A. Vainshtein, \Journal{\PLB}{293}{430}{1992};
{\it ibidem} {\bf 297}, 477 (1993) (E).

\bibitem{lephfs}
LEP $b$ lifetime Group, http://www.cern.ch/LEPHFS.

\bibitem{braun} P. Ball and V.M. Braun, \Journal{\PRD}{49}{2472}{1994}.

\bibitem{lorolam}
P. Colangelo, C.A. Dominguez, G. Nardulli and N. Paver,
\Journal{\PRD}{54}{4622}{1996}.

\bibitem{wise}  
A.V. Manohar and M.B. Wise, \Journal{\PRD}{49}{1310}{1994}.
 
\bibitem{bigi2} 
I. Bigi, preprint UND-HEP-95-BIG02 (June 1995).

\bibitem{kroll}
I.J. Kroll, preprint Fermilab-Conf-96-032,  Proceedings of the 
{\em XVII International Symposium on Lepton-Photon Interactions}, Beijing,
10-15  August 1995.

\bibitem{sachrajda}
M. Neubert and C.T. Sachrajda, \Journal{\NPB}{483}{339}{1997}.

\bibitem{bilic}
B. Guberina, R. R\"uckl  and J. Trampetic,
\Journal{\ZPC}{33}{297}{1986}.


\bibitem{der}
A. De Rujula, H. Georgi and S. Glashow, \Journal{\PRD}{12}{147}{1975};\\
J.L. Cortes and J. Sanchez-Guillen, \Journal{\PRD}{24}{2982}{1981}.


\bibitem{rev}
For a review see: B. Blok and M.A. Shifman, Proceedings of the Third
Workshop
on the Tau-Charm Factory, eds. J. Kirkby and R. Kirkby, Ed. Frontieres,
1994, pag. 269.

\bibitem{rosner}
J.L. Rosner, \Journal{\PLB}{379}{267}{1996}.

\bibitem{delphi}
P. Abreu {\em et al}., \Journal{\ZPC}{68}{375}{1995}.

\bibitem{noilam}
P. Colangelo and F. De Fazio, \Journal{\PLB}{387}{371}{1996}.

\bibitem{cinesi}C-S. Huang, C. Liu and S-L. Zhu,
\Journal{\PRD}{61}{054004}{2000}.

\bibitem{lattice4q}
M. Di Pierro {\em et al}., UKQCD Collab., hep-lat/9906031.

\bibitem{babarrad}B. Aubert {\em et al}., BaBar Collab., hep-ex/0008055.

\bibitem{bsw}
M. Bauer, B. Stech and M. Wirbel, \Journal{\ZPC}{29}{637}{1985};
\Journal{\ZPC}{34}{103}{1987}.

\bibitem{stech}
M. Neubert and B. Stech, in  {\em Heavy Flavours II},  ed.  A.J.
Buras and M. Lindner (World Scientific, Singapore) p. 294;\\
A.J. Buras and L. Silvestrini, \Journal{\NPB}{548}{293}{1999}.

\bibitem{buchalla}
M. Beneke, G. Buchalla, M. Neubert and  C.T. Sachrajda,
\Journal{\PRL}{83}{1914}{1999}; hep-ph/0006124. 

\bibitem{babar} For a comprehensive review on CP violation, see e.g. {\it
The BaBar Physics Book: Physics at an asymmetric $B$ Factory}, ed. by  
P. Harrison and H. Quinn.

\bibitem{report}P. Ball {\it et al.}, hep-ph/0003238, published in the
Proceedings of {CERN Workshop On Standard Model Physics (And More) At The 
LHC}, eds. G. Altarelli and M.L. Mangano. 

\bibitem{sk}Y. Fukuda {\em et al}., SuperKamiokande Collab.,
\Journal{\PRL}{82}{2644}{1999}.

\bibitem{eps} A. Alavi-Harati {\em et al}., KTeV Collab.,
\Journal{\PRL}{83}{22}{1999};\\
 V. Fanti {\em et al}., NA48 Collab., \Journal{\PLB}{465}{335}{1999}.

\bibitem{donut}
The DONUT Collab., announcement can be found
at www.fnal.gov/pub/donut.html. 

\end{thebibliography}
\end{document}